\input harvmac
\input epsf

\input labeldefs.tmp
\writedefs

\def\cO{{\cal {O}}}
\def\ext{{\rm Ext}}
\def\hom{{\rm Hom}}


\Title{ \vbox{\baselineskip12pt \rightline{HUTP-02/A062}
 \rightline{hep-th/0212021}
 \vskip -.25in} }{ \vbox{
\centerline{Large Volume Perspective on Branes at Singularities}
\vskip 0.1in
\centerline{} } }

\centerline{ Martijn Wijnholt}
\vskip .1in
\centerline{\sl Jefferson Laboratory of Physics, Harvard
University}
\centerline{\sl Cambridge, MA 02138, USA}
\centerline{\it  }

\vskip .1in

\vskip 0.3in

\centerline{\bf Abstract}

\noindent
In this paper we consider a somewhat unconventional approach for
deriving worldvolume theories for D3 branes probing Calabi-Yau singularities.
The strategy consists of extrapolating the calculation of F-terms
to the large volume limit. This method circumvents the inherent limitations
of more traditional approaches used for orbifold and toric singularities.
We illustrate its usefulness by
deriving quiver theories for D3 branes probing singularities where
a Del Pezzo surface containing four, five or six exceptional curves
collapses to zero size. In the latter two cases the superpotential
depends explicitly on complex structure parameters.
These are examples of probe theories for singularities which can
currently not be computed by other means.

\Date{December 2002}


\lref\CachazoSG{
F.~Cachazo, B.~Fiol, K.~A.~Intriligator, S.~Katz and C.~Vafa,
``A geometric unification of dualities,''
Nucl.\ Phys.\ B {\bf 628}, 3 (2002)
[arXiv:hep-th/0110028].
}

\lref\Karpov{
Boris V.~Karpov, Dmitri Yu.~Nogin,
``Three-block exceptional collections over Del Pezzo surfaces,''
Izvestiya Math ???
[arXiv:alg-geom/9703027].
}

\lref\BeasleyZP{
C.~E.~Beasley and M.~R.~Plesser,
``Toric duality is Seiberg duality,''
JHEP {\bf 0112}, 001 (2001)
[arXiv:hep-th/0109053].
}

\lref\FengBN{
B.~Feng, A.~Hanany, Y.~H.~He and A.~M.~Uranga,
``Toric duality as Seiberg duality and brane diamonds,''
JHEP {\bf 0112}, 035 (2001)
[arXiv:hep-th/0109063].
\hfill\break
B.~Feng, A.~Hanany and Y.~H.~He,
``Phase structure
of D-brane gauge theories and toric duality,'' JHEP {\bf 0108},
040 (2001) [arXiv:hep-th/0104259].
\hfill\break
B.~Feng, S.~Franco, A.~Hanany and Y.~H.~He,
``Unhiggsing the del Pezzo,'' arXiv:hep-th/0209228.
\hfill\break
A.~Hanany and A.~Iqbal,
``Quiver theories from
D6-branes via mirror symmetry,'' JHEP {\bf 0204}, 009 (2002)
[arXiv:hep-th/0108137].
}

\lref\Macaulay{
Daniel R.~Grayson and Michael E.~Stillman,
``Macaulay 2, a software system for research in algebraic geometry,''
Available at http://www.math.uiuc.edu/Macaulay2/
}

\lref\Smith{
Gregory G.~Smith,
``Computing Global Extension Modules for Coherent Sheaves on a Projective
   Scheme,''
Journal of Symbolic Computation {\bf 29}  (2000) 729-746
[arXiv:math.AG/9807170].
}

\lref\DouglasGI{
M.~R.~Douglas,
``D-branes, categories and N = 1 supersymmetry,''
J.\ Math.\ Phys.\  {\bf 42}, 2818 (2001)
[arXiv:hep-th/0011017].
\hfill\break
I.~Brunner, M.~R.~Douglas, A.~E.~Lawrence and C.~Romelsberger,
``D-branes on the quintic,''
JHEP {\bf 0008}, 015 (2000)
[arXiv:hep-th/9906200].
\hfill\break
D.~Berenstein and M.~R.~Douglas,
``Seiberg duality for quiver gauge theories,''
arXiv:hep-th/0207027.
}

\lref\KatzGH{
S.~Katz and E.~Sharpe,
``D-branes, open string vertex operators, and Ext groups,''
arXiv:hep-th/0208104.
}

\lref\Bondal{
A.~Bondal and D.~Orlov,
``Semiorthogonal decomposition for algebraic varieties,''
arXiv:alg-geom/9506012.
}

\lref\DouglasFR{
M.~R.~Douglas, S.~Govindarajan, T.~Jayaraman and A.~Tomasiello,
``D-branes on Calabi-Yau manifolds and superpotentials,''
arXiv:hep-th/0203173.
}

\lref\DouglasAH{
M.~R.~Douglas, B.~Fiol and C.~Romelsberger,
``Stability and BPS branes,''
arXiv:hep-th/0002037.
}

\lref\DouglasQW{
M.~R.~Douglas, B.~Fiol and C.~Romelsberger,
``The spectrum of BPS branes on a noncompact Calabi-Yau,''
arXiv:hep-th/0003263.
}

\lref\PolchinskiMT{
J.~Polchinski,
``Dirichlet-Branes and Ramond-Ramond Charges,''
Phys.\ Rev.\ Lett.\  {\bf 75}, 4724 (1995)
[arXiv:hep-th/9510017].
}

\lref\DouglasSW{
M.~R.~Douglas and G.~W.~Moore,
``D-branes, Quivers, and ALE Instantons,''
arXiv:hep-th/9603167.
}

\lref\MorrisonCS{
D.~R.~Morrison and M.~R.~Plesser,
``Non-spherical horizons. I,''
Adv.\ Theor.\ Math.\ Phys.\  {\bf 3}, 1 (1999)
[arXiv:hep-th/9810201].
}

\lref\WittenFB{
E.~Witten,
``Chern-Simons gauge theory as a string theory,''
Prog.\ Math.\  {\bf 133}, 637 (1995)
[arXiv:hep-th/9207094].
}

\lref\MayrAS{
P.~Mayr,
``Phases of supersymmetric D-branes on Kaehler manifolds and the McKay  correspondence,''
JHEP {\bf 0101}, 018 (2001)
[arXiv:hep-th/0010223].
}

\lref\HoriCK{
K.~Hori, A.~Iqbal and C.~Vafa,
``D-branes and mirror symmetry,''
arXiv:hep-th/0005247.
}

\lref\BershadskyCX{
M.~Bershadsky, S.~Cecotti, H.~Ooguri and C.~Vafa,
``Kodaira-Spencer theory of gravity and exact results for quantum string amplitudes,''
Commun.\ Math.\ Phys.\  {\bf 165}, 311 (1994)
[arXiv:hep-th/9309140].
}

\lref\inprogress{F.~Cachazo, S.~Katz, M.~Ro\~cek, C.~Vafa, M.~Wijnholt.}

\lref\LawrenceJA{
A.~E.~Lawrence, N.~Nekrasov and C.~Vafa,
``On conformal field theories in four dimensions,''
Nucl.\ Phys.\ B {\bf 533}, 199 (1998)
[arXiv:hep-th/9803015].
}

\lref\IbanezQP{
L.~E.~Ibanez, R.~Rabadan and A.~M.~Uranga,
Nucl.\ Phys.\ B {\bf 542}, 112 (1999)
[arXiv:hep-th/9808139].
}

\lref\TomasielloYM{
A.~Tomasiello,
``D-branes on Calabi-Yau manifolds and helices,''
JHEP {\bf 0102}, 008 (2001)
[arXiv:hep-th/0010217].
}

\lref\GovindarajanVI{
S.~Govindarajan and T.~Jayaraman,
``D-branes, exceptional sheaves and quivers on Calabi-Yau manifolds: From  Mukai to McKay,''
Nucl.\ Phys.\ B {\bf 600}, 457 (2001)
[arXiv:hep-th/0010196].
}

\lref\FengKK{
B.~Feng, A.~Hanany, Y.~H.~He and A.~Iqbal,
``Quiver theories, soliton spectra and Picard-Lefschetz transformations,''
JHEP {\bf 0302}, 056 (2003)
[arXiv:hep-th/0206152].
}

\lref\HerzogDJ{ C.~P.~Herzog and J.~Walcher,
``Dibaryons from
exceptional collections,'' JHEP {\bf 0309}, 060 (2003)
[arXiv:hep-th/0306298].
}

\lref\HerzogWT{
C.~P.~Herzog and J.~McKernan,
``Dibaryon spectroscopy,''
JHEP {\bf 0308}, 054 (2003)
[arXiv:hep-th/0305048].
}

\listtoc

\writetoc

\newsec{Overview}

\bigbreak

The description of certain extended objects in string theory as
Dirichlet branes \PolchinskiMT\ has given us a new window into the physics of highly
curved geometries. The worldvolume theory of a D-brane probing an orbifold
singularity at low energies is described by a gauge theory of
quiver type \DouglasSW. The correspondence goes both ways; on the one hand,
we can access distance scales that are shorter than the scales we
can probe with closed strings. On the other hand, one may view
this as a way of engineering interesting gauge theories in string
theory, and then using string dualities we can sometimes learn new
things about gauge theories. Let us focus on  D3 branes probing
Calabi-Yau three-fold singularities, which give rise to ${\cal N} = 1$ gauge
theories in 3+1 dimensions.

Orbifolds of flat space constitute an interesting and
computationally convenient set of singular space-times, and have given us
a nice picture of the probe brane splitting up into fractional branes
near the singularity\foot{
These are the analogues of wrapped branes when we resolve the
singularity and go to large volume.}.
But they
are certainly not the most general type of singularity and do not
lead to the most general allowed gauge theory.
Another  class is that of toric
singularities, which contains the class of abelian orbifold
singularities. These backgrounds are described by toric geometry (or linear sigma
models). Adding some probe branes
filling the remaining transverse dimensions leads to SUSY quiver theories
where the ranks of the gauge groups are all equal and the
superpotential can be written in such a way that no matter field
appears more than twice. The prime example of this type of
singularity is the conifold.

The way one usually analyses toric singularities is by embedding
them
 in an orbifold singularity, which we
know how to deal with \MorrisonCS. Then one may partially resolve the
singularity in a toric way to get the desired space. In the gauge
theory this corresponds to giving large VEVs to certain fields and
integrating out the very massive modes. At sufficiently low energies
the gauge theory only describes the local neighbourhood of the toric
singularity.

This however does not yet exhaust the class of allowed Calabi-Yau
singularities or ${\cal N} = 1$ gauge theories.
We refer to the remaining cases as non-toric
singularities.

In this article we would like to attack these cases by using an
approach different from  the ones we have mentioned above. The
strategy consists of extrapolating to large volume where the
correct description of the fractional branes is in terms of
certain collections of sheaves\foot{
For our purposes these will mostly be ordinary vector bundles,
even line bundles.},
called exceptional collections. The matter content and
superpotential can be completely determined by geometric
calculations involving these collections, and the results are not
affected by the extrapolation. In principle this yields a general approach to
all Calabi-Yau singularities, both toric and non-toric.

In the next section we will review many of the ingredients of this
approach\foot{
Some references:
\refs{\WittenFB,\BershadskyCX,\HoriCK,\CachazoSG,\DouglasGI,\DouglasQW}}.
In the following section we will apply the techniques to compute the
quiver theories for branes probing certain Calabi-Yau geometries where a
four-cycle collapses to zero size. If the singularity can be resolved
by a single blow-up, such a four-cycle is
either a ${\bf P}^1$ fibration over a Riemann surface or a Del Pezzo surface.
We will treat some toric and non-toric Del Pezzo cases in detail,
computing their moduli spaces and  showing how they are related to
known quiver theories through the  Higgs mechanism.
The non-toric
cases considered here are new and there is  no other method we know of
for calculating them. The case of Del Pezzo 6 in particular can be viewed as a
four parameter non-toric family of deformations of the familiar
${\bf C}^3/Z_3 \times Z_3$ orbifold singularity.

In order to do the computations we introduce the notion of a `three-block' quiver diagram
in section three. Quiver diagrams of this type have some important
simplifying features. Most importantly for our purposes is that
they they lead to superpotentials with only cubic terms for all
the Del Pezzo cases. Other quiver theories may then be deduced
through Seiberg duality.

\newsec{Large volume perspective}

\subsec{Relation to exceptional collections}

We will start with a non-compact Calabi-Yau threefold which has a
4-cycle or 2-cycle shrinking to zero size at a singularity. Then we
may put a set of D3 branes at the singularity and try to understand
the low energy worldvolume theory on the D3
branes, which turns out to be an ${\cal N} = 1$ quiver gauge theory.
The analysis may be done using exceptional collections living
on the shrinking cycle.

We recall first the definition of an exceptional collection.
An exceptional sheaf  is a sheaf $E$ such that dim ${\rm Hom}(E,E) = 1$
and ${\rm Ext^k(E,E)}$ is zero whenever $k>0$. An exceptional
collection is an ordered collection $\{E_i\}$  of exceptional sheaves
such that ${\rm Ext}^k(E_i,E_j) = 0$ for any $k$ whenever $i>j$. We
will usually assume an exceptional collection to be complete in the
sense that the Chern characters of the sheaves in the collection
generate all the cohomologies of the shrinking cycle
{\foot{
A more precise definition would probably be that the collection
generates the bounded derived category of coherent sheaves.}}.

A sheaf living on a cycle inside the
Calabi-Yau gives rise to a sheaf on the Calabi-Yau by
``push-forward.'' Heuristically this just means that we take the sheaf
on the cycle and extend it by zero to get a sheaf denoted by $i_* E$
on the CY. An exceptional collection lifts to a set of sheaves with
``spherical'' cohomology on the Calabi-Yau: because of the formula
\eqn\aa{
 {\rm Ext}^k(i_* E_i, i_* E_j)\sim \sum_{l+m = k}
{\rm Ext}^{l}(E_i,E_j \otimes \Lambda^m N)   }
where $N$ is the normal bundle to the cycle we
conclude that the cohomologies of an exceptional
 sheaf are given by dim ${\rm Ext}^k(i_*
E,i_* E) = \{ 1,0,0,1\}$ for $k = \{0,1,2,3\}$. The reason for the
terminology is that under mirror symmetry
such sheaves get mapped to Lagrangian 3-spheres. The ordering of
the sheaves is related to the ordering with respect to the
imaginary coordinate on the $W$-plane.

%
%
\bigskip
\centerline{\epsfxsize=.5\hsize\epsfbox{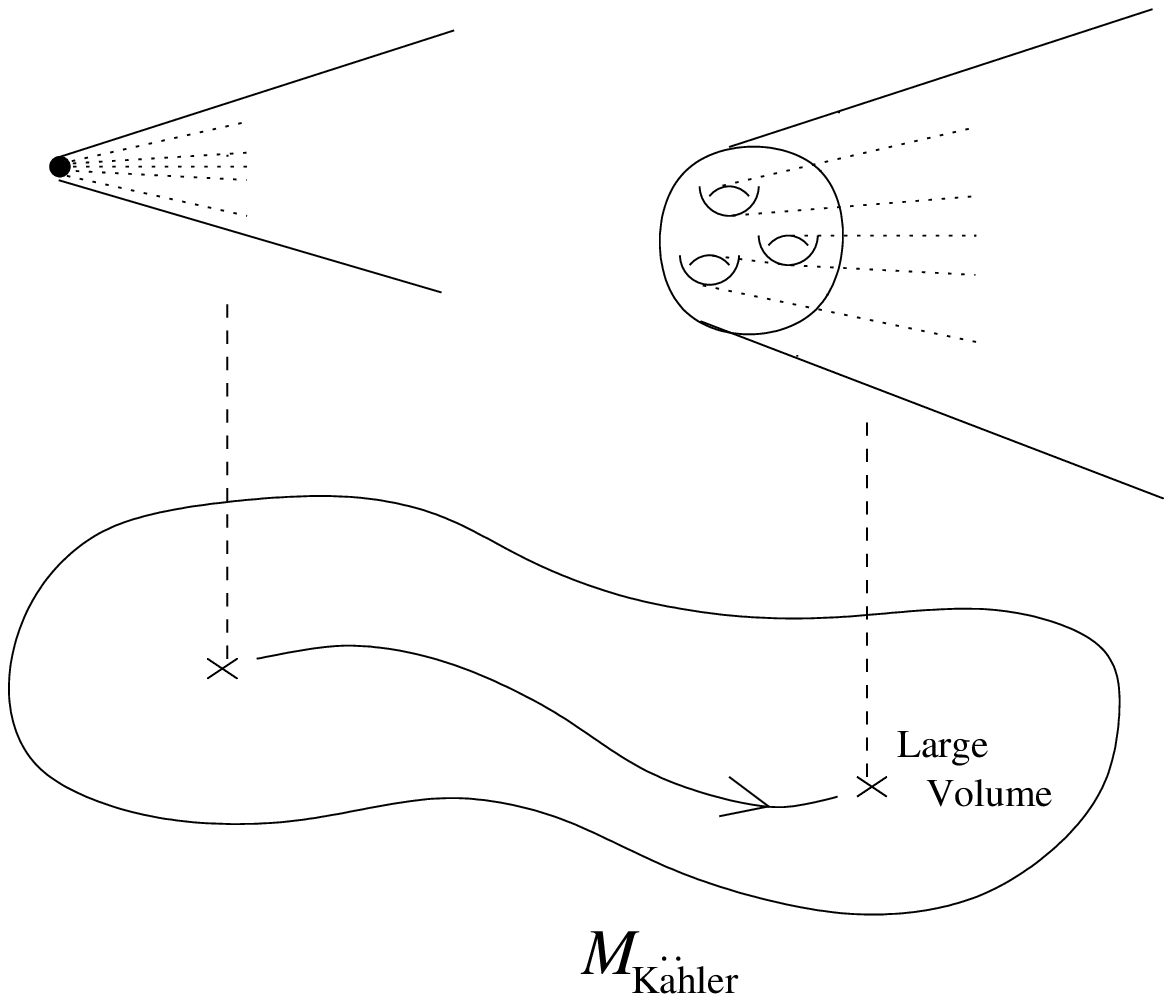}}
\nobreak\noindent{\ninepoint\sl \baselineskip=8pt \fig\largevolume{}:
{\sl Extrapolation to large volume in order to perform calculations.}}
\bigskip


The significance of exceptional collections is that the sheaves in the
collection have the right properties to be
the large volume descriptions of the fractional branes
at the singularity. Since we are in type IIB string theory, D-branes
filling the 3+1 flat directions must wrap even dimensional cycles in
the CY, and therefore strings ending on such branes satisfy boundary
conditions that allow for a B-type topological twist. Correlation
functions
in the twisted theory are independent of the choice of
K\"ahler structure on the Calabi-Yau, and correspond to
superpotential terms in the untwisted theory. The upshot is that K\" ahler
parameters only manifest themselves in the D-terms of the gauge
theory. So as long as we are asking questions only about the F-terms
(superpotential terms), we can do a K\"ahler deformation to give the
vanishing cycle a finite volume. At large volume we can describe
D-branes as sheaves and the computations can be done using exceptional
collections. It is important to remember that the gauge theory
we will be discussing doesn't
actually live at large volume, but using the justification above
we will interpret the results of
the geometric computations as protected quantities in a gauge theory
which is a valid low energy description of the D3 branes at small volume.

\subsec{Spectrum and quiver diagram}

In order to describe the gauge theory on the D3 branes we need to
construct the D3 brane out of the fractional branes and find the
lightest modes. A D3 brane filling the 3+1 flat directions is
determined by specifying
a point $p$ on the Calabi-Yau, so we will use a skyscraper sheaf
$\cO_p$ to represent it. The RR charges of a brane are combined
in the Chern character of a sheaf, so if $\{ E_i \}$ is an exceptional
collection and $n_i$ the multiplicities of the fractional branes then
we have the condition:
\eqn\sumcharges{ \sum_i n_i\, {\rm ch}(i_* E_i) = {\rm ch} (\cO_p) .}
Notice that some of the $n_i$ are necessarily negative, because we
have to cancel all the charges associated with wrappings of 2- and
4-cycles. So it may appear that we have both ``branes'' and
``anti-branes'' present and therefore break supersymmetry, however
this is just an artefact of the large volume description. As we vary
the K\"ahler moduli to make the Calabi-Yau singular the central charges\foot{
The central charge is associated with a particle, not with a
space-filling brane, but the two systems are closely related as far as
SUSY properties is concerned so we will borrow the language.}
of the fractional branes all line
up and thus they all break the same half of the supersymmetry at the
point of interest\foot{
A nice picture of this for the case of ${\bf C}^3/Z_3$ can be found in
\DouglasQW.}.

Now the lightest modes for strings with both endpoints on
the same fractional brane fit in an ${\cal N} = 1$
vector multiplet. Since the fractional brane is rigid ($\ext^1(E,E) =
0$ from the sheaf point of view) we do not get any adjoint chiral multiplets
describing deformations from these strings. If the fractional
brane has multiplicity $| n_i |$ then the vector multiplet will
transform in the adjoint of $U(| n_i |)$.

We can also have strings with
endpoints on different fractional branes $E_i$ and $E_j$.
The lightest modes of these
strings are ${\cal N} = 1$ chiral fields transforming in the
bifundamental of the gauge groups associated with the two fractional
branes. To count their number we can do a computation in the B-model
which we can perform at large volume. If the branes fill the whole
Calabi-Yau then the ground states of these
strings can  be seen to arise from the cohomology of the Dolbeault
operator coupled to the gauge fields of the two fractional branes, $Q
= \bar{\del}_{\bar{z}} + A^{(j)}_{\bar{z}} - A^{(i)T}_{\bar{z}}$,
acting on the space of anti-holomorphic forms $\Omega^{(0,
\cdot)}(E_i^* \otimes E_j)$. If the branes are wrapped on lower
dimensional cycles then we can dimensionally reduce by changing the
gauge fields with indices that do not lie along the worldvolume of the
branes into normal bundle valued scalars. Schematically then
the number of chiral fields is in one to one correspondence with the
generators of the sheaf cohomology groups $H^{(0,m)}(E_i^* \otimes E_j \otimes \Lambda^n
N)$ where $N$ is the normal bundle of the shrinking cycle that both
branes are wrapped on, or more
generally
the global Ext groups ${\rm Ext}^k(i_* E_i,i_* E_j)$ \KatzGH. However we are double counting
because given a generator
we can get another one by applying Serre duality on the three-fold. As we discuss momentarily
the corresponding
physical mode would have opposite charges and opposite chirality, so this
should be
part of the vertex operator for the corresponding anti-particle. Then we should have
only one chiral field for each pair of generators that are related by
Serre duality.
Finally the degree of the Ext group indicates  ghost number $k$
 of the topological vertex operator. In the physical theory
this gets related to the chirality of the multiplet through the GSO projection.
If the degree is even we
get say a left handed fermion
so we should assign fundamental charges for
the $j$th gauge group and anti-fundamental charges of the $i$th gauge
group to the chiral field. Then if $k$ is odd we get a right handed fermion
and we should assign the opposite charges to the chiral field.
The chirality flips if we turn a brane the
string ends on into an anti-brane (or more precisely if we invert its
central charge) because this shifts $k$ by an odd integer.

We may summarise this spectrum in a quiver diagram. For each
fractional brane $E_i$ we draw a node which corresponds to a vector
multiplet transforming in the adjoint of $U(|n_i|)$ (or $U(|n_iN|)$
if we started with $N$ D3 branes). For each chiral multiplet in the
 fundamental of $U(|n_j|)$ and in the anti-fundamental of
$U(|n_i|)$ we draw an arrow from node $i$ to node $j$. Let us
illustrate this in the example of ${\bf C}^3/Z_3$ where we can do a K\"ahler
deformation to get a finite size ${\bf P}^2$ (this example was treated
in \refs{\CachazoSG,\DouglasQW}). We'll take the exceptional collection on ${\bf P}^2$ to be
the set  $\{ \cO(-1),\Omega^1(1),\cO(0) \}$. By $\Omega^1(1)$ we mean
the cotangent bundle, tensored with the line bundle $\cO(1)$. To reproduce
the charge of a D3 brane the multiplicities will have to be
$\{ 1,-1,1 \}$. Then we will draw three nodes with a gauge group
 $U(N)$ for each. The
non-zero Ext's are found to be
\eqn\aa{ {\rm dim}\ {\rm Ext}^0(E_1,E_2) = 3, \quad  {\rm dim}\ {\rm Ext^0}(E_2,E_3) = 3,
\quad  {\rm dim}\ {\rm Ext}^0(E_1,E_3) = 3 }
with everything else either zero or related by Serre duality.
Drawing in the arrows with the proper orientations gives the following
quiver diagram:
\bigskip
\centerline{\epsfxsize=.64\hsize\epsfbox{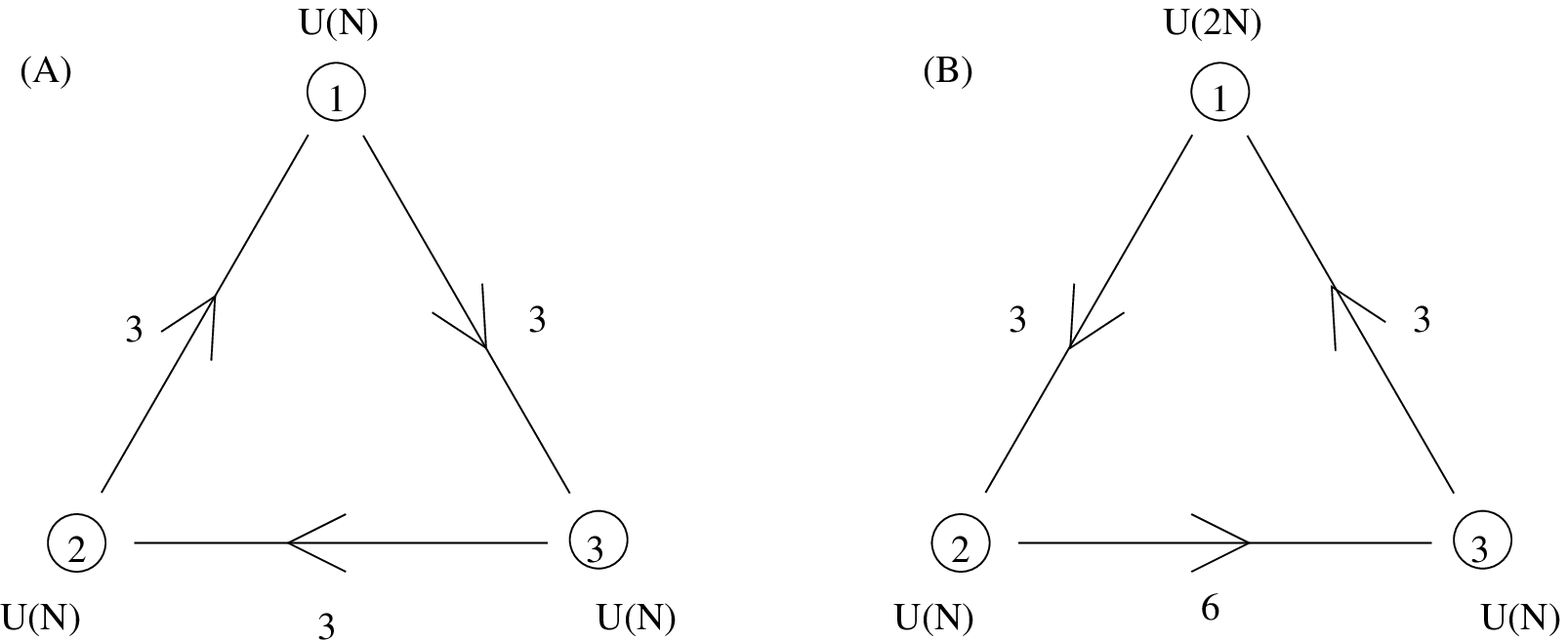}}
\noindent{\ninepoint\sl \baselineskip=8pt \fig\Ptwo{}:
{\sl (A): Quiver for the orbifold ${\bf C}^3/Z_3$. (B): Seiberg dual
quiver, after integrating out massive modes.}}
\bigskip

If one is only interested in the net number of arrows between two
nodes, it is sufficient to find the relative Euler character of the
two corresponding sheaves
\eqn\aa{ \chi(i_* E,i_* F) = \sum_k (-1)^k {\rm dim}\  \ext^k(i_* E, i_* F).}
This number is frequently easier to compute than the actual number of
arrows.\foot{
For instance for the case of exceptional sheaves $E$ and $F$ on Del Pezzo
surfaces that we
will discuss later on, the number of arrows is simply given by
$r_E \, d_F - r_F \, d_E$ where $r$ is the rank of the
sheaf and $d$ is the degree (i.e. the intersection with the canonical class $K$).}
Now notice that  a brane
can only have non-zero intersection number with something that is
localised at a point on the Calabi-Yau if it fills the whole
Calabi-Yau.
This does not happen for fractional branes, which are localised
around the cycle that shrinks to zero. Hence
we deduce that
\eqn\anomcancel{
    \chi(\cO_p,i_*E_j) = \sum_i n_{ij} n_i = 0 }
where $n_{ij}$ is the net number of arrows from $E_i$ to $E_j$.
In terms of field theory
equation \anomcancel\ states that the number of arrows directed
towards node $j$ exactly matches the number of arrows pointing away
from it, or that the number of quarks for the $j$th gauge group is the
same as the number of anti-quarks. We are therefore guaranteed that
gauge anomalies cancel \CachazoSG.\foot{ At least for the
non-abelian factors; for the $U(1)$'s there's mixing with closed string modes, see \IbanezQP.}

\subsec{Superpotential}

So far we have only indicated how to obtain the matter content of the
gauge theory, but we know for instance from the ${\bf C}^3/Z_3$
orbifold that we should also expect a superpotential. Another way
to see this is by observing that the moduli space of the gauge theory
should be the moduli space of the sheaf $\cO_p$, which certainly
contains the
Calabi-Yau three-fold itself.
We can only recover this by imposing  additional constraints from a
superpotential on the VEVs. As we have discussed above, the
superpotential is also a piece of data we can compute at large volume.

The cubic terms in the superpotential can be computed with relative
ease. Suppose we want to know if a cubic coupling for the chiral
fields running between nodes $i$,$j$ and $k$. Then after
picking Ext generators
which correspond to the chiral fields we may compute the Yoneda
pairings
\eqn\aa{  \ext^l(i_* E_i,i_*E_j) \times
\ext^m(i_* E_j,i_*E_k) \times \ext^n(i_* E_k,i_*E_i) \to
\ext^{l+m+n}(i_* E_i,i_*E_i) .}
If $l+m+n = 3$ then we may use the fact that ${\rm dim} \ \ext^3(E,E)=
1$
on a CY three-fold to get a number from the above composition. This number is
the coefficient of the cubic coupling in the superpotential. If
$l+m+n \not = 3$ then the cubic coupling is automatically zero. One
may justify this claim  by looking at a disc diagram with three vertex
operators inserted at the boundary. Eg. the cubic part of a superpotential term in the
action of the form
\eqn\aa{
{\rm Tr}\left({\del^2  W(\phi)\over \del\phi^2} \psi \psi\right) }
is computed in string theory by a disc diagram with two fermion vertex
operators and one scalar vertex operator on the boundary. This is just
the type of diagram that can be computed in the topologically twisted
theory
\eqn\threepoint{
{\rm Tr}\ \vev{V_\phi V_\psi V_\psi}_{\rm untwisted} \sim
\vev{V_{NS}^{(1)}  V_{NS}^{(2)} V_{NS}^{(3)}}_{\rm twisted}
}
where the $V^{(i)}$ are the internal part of the full physical vertex
operators, with some spectral flow applied if the internal part was in the RR
sector. At large volume the  $V^{(i)}_{NS}$ can be represented as generators of the sheaf
cohomology from which the physical fields descended. In the B-model
the amplitude \threepoint\ just gives the overlap of the vertex operators and
vanishes unless the total ghost number from the three vertex operators
adds up to three.

For ${ \bf P}^2$  one finds the following generators \refs{\CachazoSG,\DouglasQW}:
\eqn\sections{\eqalign{
(C_3 z_2 - C_2 z_3)dz_1 + (C_1 z_3 - C_3 z_1) dz_2 + (C_2 z_1 - C_1 z_2)
dz_3  &\in  \hom({\cO} (-1),\Omega^1(1)) ,\cr
A_1 {\del\over \del z_1} + A_2 {\del\over \del z_2} + A_3  {\del\over \del z_3}
& \in  \hom(\Omega^1(1), {\cO}), \cr
B_1 z_1^* + B_2 z_2^* + B_3 z_3^* & \in  \hom^*({\cO} (-1),{\cO}). }}
The star indicates we have dualised the Hom, making it an
$\ext^3$ on the CY by Serre duality.
 Computing the Yoneda pairings gives the usual orbifold
superpotential
\eqn\Bo{ W = {\rm Tr}\left(A_1 (B_2 C_3 - B_3 C_2) + A_2 (B_3 C_1 -
B_1 C_3)
+ A_3 (B_1
C_2 - B_2 C_1) \right)}

It turns out that for a clever choice of exceptional collections for
Del Pezzo surfaces, namely the ones with ``three-block'' structure
\Karpov, the superpotential only has cubic
couplings. So for our purposes
we do not have to worry about higher order terms in
the superpotential. However let us briefly discuss how one might go
about computing the higher order terms \CachazoSG.\foot{
Apart from a direct computation in holomorphic Chern-Simons theory \WittenFB,
which is usually quite hard.}
Suppose we are
interested in the coefficient of a possible quartic term, say
involving nodes $i,j,k$ and $l$. Then we would like to compute an
amplitude of the form
\eqn\fourpoint{
\vev{V_{NS}^{(1)} V_{NS}^{(2)} V_{NS}^{(3)} \int V_{NS}^{(4)}}   }
Here we have assumed that the operator $V_{NS}^{(4)}$ comes from an
$\ext^1$, so its ghost number is equal to 1 and can be integrated over
the boundary between the points where $V_{NS}^{(3)} $ and
$V_{NS}^{(1)}$  are inserted. Suppose that instead we consider the
following amplitude:
\eqn\aa{  \vev{ V_{NS}^{(1)} V_{NS}^{(2)} V_{NS}^{(3)} \exp\left[
t \int V_{NS}^{(4)} \right]}
=  \vev{ \tilde{V}_{NS}^{(1)} V_{NS}^{(2)} \tilde{V}_{NS}^{(3)} }_t
}
In other words we use the operator $V_{NS}^{(4)} $
 to deform the sigma model action. Then we may recover \fourpoint\ by
 differentiating with respect to $t$. If $V_{NS}^{(4)}$
 comes from $\ext^1(E_i,E_j)$ then this deformation creates a new
 boundary condition corresponding to the sheaf $F$, where $F$ is the
 deformation of   $E_i \oplus E_j$ defined by the extension class.
In gauge theory terms we are Higgsing down the quartic term to get a
 cubic term. Now we may proceed as before and compute a cubic coupling using $F,
E_k$ and $E_l$.\foot{
In general when we have  $\ext^n$ with $n\not = 1$, there still exists
a deformation but it
does not give a sheaf but a complex of sheaves;
it should still be possible to do
some  calculation in this case. In the disc diagram we should have
SL(2,R) invariance so it seems somewhat peculiar that the $\ext^1$'s
are singled out.}

Finally to completely specify the gauge theory we must also supply
the K\" ahler terms. We will not have anything to say about this,
but one should keep in mind that we are considering the far IR
physics which is expected to give rise to an interacting
superconformal theory based on ADS/CFT arguments \MorrisonCS.
In this case the K\"ahler terms are quite possibly
completely fixed in terms of F-term
data by superconformal invariance, as they are thought to be for
two-dimensional superconformal theories.

\subsec{Mutations and Seiberg duality}

In our discussion up till now we have suppressed the fact that there actually are many
choices of exceptional collections. Indeed there is a general procedure
for making new exceptional collections from old ones, called
mutation. At the level of Chern characters this transformation looks very familiar.
Given an exceptional collection $\{ \ldots ,E_{i-1},E_i,E_{i+1}, E_{i+2},
\ldots \}$ we can make a new "left mutated" collection\foot{
There is also a right shifted version
$\{ \ldots ,E_{i},R_{E_i}E_{i-1},E_{i+1},\ldots \}$.}
\eqn\aa{\{ \ldots ,E_{i-1},L_{E_i}E_{i+1},E_{i},E_{i+2}, \ldots \} }
where the Chern character of the new sheaf
$L_{E_i}E_{i+1}$ is given by\foot{
When we apply this formula in the next section we will deduce the
sign using charge conservation.}
\eqn\monodromy{
{\rm ch}(L_{E_i}E_{i+1}) = \pm \left[{\rm ch }(E_{i+1}) - \chi(E_{i},E_{i+1})
{\rm ch}(E_i)\right] .}
This looks very similar to (and is in fact mirror to)
Picard-Lefschetz monodromy. One may construct the sheaf
$L_{E_i}E_{i+1}$ out of $E_i$ and $E_{i+1}$ using certain exact
sequences.
\bigskip
\centerline{\epsfxsize=0.99\hsize\epsfbox{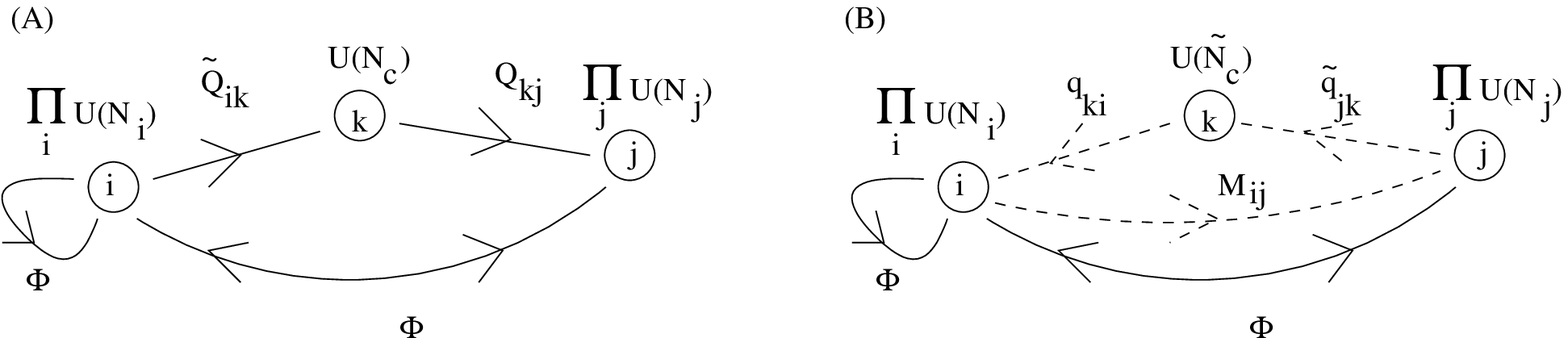}}
\noindent{\ninepoint\sl \baselineskip=8pt \fig\sdual{}:
{\sl (A): Organising the nodes before applying Seiberg duality to node
$k$. (B): Seiberg dual theory.}}
\bigskip


For a certain class of mutations the corresponding
transformation on the quiver gauge theory was interpreted as Seiberg
duality  \CachazoSG. Before we delve into this let us first comment on
the abelian factors in the gauge groups. There is an overall factor of
$U(1)$ which completely decouples and then there are relative $U(1)$'s
between the nodes which become weakly coupled in the infrared. Seiberg
duality is a statement about the infrared behaviour of the non-abelian
part of the gauge theory, which can get strongly coupled. So when we
discuss a quiver diagram with $U(N)$ gauge groups we would  like to
think of the $SU(N)$ parts as dynamical gauge groups and the $U(1)$'s
as global symmetries, or alternatively reinstate the $U(1)$'s as gauge
symmetries only at ``intermediate'' energy scales.

To see which class of mutations we need to look at
let us suppose we want to do a Seiberg duality on node $k$ and
organise the quiver so that all the incoming arrows (labelled by $i$)
are to the
left of node $k$ and all the outgoing arrows
(labelled by $j$) to the right of
node $k$ as in \sdual A.
We think $SU(N_k)$ as the colour group which gets strongly coupled
and treat all the other nodes as
flavour groups. The total number of flavours is
\eqn\aa{ N_f = \sum_i n_{ik} N_i = \sum_j n_{kj} N_j }
where the second equality comes from anomaly cancellation as we
discussed above.
In the Seiberg dual theory the quarks and anti-quarks $\{\tilde{Q}_{ik}, Q_{kj} \}$
are replaced by dual quarks  $\{q_{ki},\tilde{q}_{jk} \}$,
which means we have to reverse the arrows going into and coming
out of node $k$ in the quiver diagram. We also have to add
mesons $M_{ij}$ between the nodes on the
left of $k$ and the nodes on the right of $k$. In the original theory
these mesons are bound states of the quarks, $M_{ij} \sim
\tilde{Q}_{ik} Q_{kj}$, but they become fundamental in the dual
theory.
Finally we have
to replace the original gauge group $SU(N_c)= SU(N_k)$ on node $k$ by the
dual gauge group $SU(\tilde{N}_c) = SU(N_f - N_c)$.
If the superpotential for the original
theory is $W(\tilde{Q} \cdot Q,\Phi) $ then the superpotential for the dual
theory is
\eqn\aa{
 W_{\rm dual}(\tilde{q} \cdot q,M,\Phi)
 = W_{\rm orig}(M,\Phi) + \lambda\  {\rm Tr}(M\tilde{q} q) .}
Since we are only interested in F-terms, the coupling constant
$\lambda$ is not relevant for our purposes and can be scaled away.

This transformation can be realised in geometry
by  doing a mutation by $E_k$ on either (1) all the sheaves $E_i$
to the left of
$k$ or (2) all the sheaves $E_j$ to the right of $k$ \CachazoSG.\foot{
Some progress has been made towards the understanding of more general
mutations \inprogress,
though as we will note in section 3 one doesn't necessarily expect an
arbitrary mutation to give rise to a good field theory.}
Let us illustrate this for the case of ${\bf P}^2$. Starting with the
collection $\{ \cO(-1),\Omega^1(1),\cO(0) \}$ we may do a Seiberg
duality on node 1 by going to the mutated collection
\eqn\aa{ \{L_{\cO(-1)} \Omega^1(1), \cO(-1),\cO(0) \}
 =  \{ \cO(-2), \cO(-1),\cO(0) \} .}
Note that in \Ptwo B we label the nodes to be consistent with the
original quiver \Ptwo A, not according to the order of the exceptional
collection. Now for this new collection
 the matter fields descend from monomials on  ${\bf P}^2$:
\eqn\aa{ \eqalign{
A_i z^i & \in \hom(\cO(-2),\cO(-1)) \cr
B_j z^j & \in \hom(\cO(-1),\cO(0)) \cr
C^*_{ij}z^i  z^j & \in \hom(\cO(-2),\cO(0)) .}}
Here  $C_{ij}$ is symmetric in $i$ and $j$, so there are only six
fields, not nine. From the Yoneda pairings we deduce the superpotential
\eqn\aa{ W = {\rm Tr} \ \sum_{i,j} C_{ij} A_i B_j }
An easy calculation shows that this agrees with the Seiberg dual of
\Ptwo A after integrating out the massive fields, which are irrelevant in
the infrared. We create nine new mesons between nodes 2 and 3 but
three of these mesons pair up with fields from the original theory, leaving
six massless fields.

A crucial property of Seiberg duality is that it preserves the
moduli space and the ring of chiral operators of the theory. The
moduli space is parametrised by the gauge invariant operators
modulo algebraic relations and relations coming from the
superpotential. We may distinguish between the mesonic operators
which are products of "quarks" and "anti-quarks" (i.e. they are
gauge invariant because fundamentals are tensored with
anti-fundamentals) and the baryonic operators which are gauge
invariant because they contain an epsilon-tensor. In this article
we will investigate the part of the moduli space parametrised by the mesonic
operators. These moduli describe the motion of the branes in the
background geometry.
In particular for a single probe brane we expect to recover
the background geometry itself, perhaps with some additional branches
if the singularity is not isolated. The baryonic operators describe partial
resolutions of the background (coming from twisted sector closed
strings in the case of orbifolds),
and we will make use of them for
comparing the new quiver theories we write down to known quivers
in the partially resolved background.

\newsec{Superpotentials for toric and non-toric Calabi-Yau singularities}

In light of the approach sketched in the previous section, the first step in
writing down the quiver theory for branes at a singularity is identifying
a suitable exceptional collection of sheaves at large volume.
The Calabi-Yau three-fold singularities
 we would like to consider here are complex cones over
Del Pezzo surfaces (though we will also make some comments on the
conifold).
This just means that we take the equations
defining a Del Pezzo surfaces living in projective space, and regard
the same equations as defining a variety over affine space, which
gives a cone over the surface with a singularity at the origin where
the 4-cycle shrinks to zero size.

Since the Del Pezzo's
can be obtained from ${\bf P}^2$ or ${\bf P}^1 \times {\bf P}^1$ by
blowing up generic points, it would be nice to have a simple way to
construct exceptional sheaves on the blow-up from exceptional sheaves
on the original variety. This is indeed possible; we can start for
instance with the exceptional collection $\{ \cO(-1),\cO(0),\cO(1) \}$
on  ${\bf P^2}$, and let $\sigma$ denote the map from the $n$th Del
Pezzo to ${\bf P^2}$ that blows down the exceptional curves $E_1$
through $E_n$. Then an
exceptional collection on the $n$th Del Pezzo is given by
$\{ \sigma^* \cO(-1),\sigma^*\cO(0),\sigma^*\cO(1), \cO_{E_1}, \ldots,
\cO_{E_n} \}$  \Bondal.\foot{
We thank R.~Thomas for pointing out this reference.}
However this collection is probably not very
useful for constructing quiver theories. The reason is that the
$\cO_{E_i}$ are needed to recover the full moduli space, but since only
$\cO_{E_i}$ carries charge for the $E_i$ cycle and such charges must
cancel, it follows that it must appear both as a brane and an
anti-brane in the quiver, thereby at the least breaking supersymmetry.

Other exceptional collections are related by mutation, so we can take
the exceptional collection mentioned above as a starting point and try
to get better behaved collections. A  list of some exceptional
collections obtained in this way was given in \Karpov. If we group all the sheaves that have no
relative cohomologies between them together in a single ``block'', then the
exceptional collections written down  in \Karpov\ only consist of three  blocks. Such collections
have many useful properties and provide the easiest way to access the Del Pezzo cases;
eg. the matter fields are always given by
$\ext^0$'s (corollary 3.4 in \Karpov) and the expected superconformal invariance of the gauge theory
restricts the
superpotential to be purely cubic. These properties simplify the computations significantly.
The allowed three-block collections were completely determined in \Karpov\ by
studying a Markov-type equation
\eqn\markov{
\alpha x^2 + \beta y^2 + \gamma z^2 = \sqrt{\alpha \beta \gamma
K^2} x y z. }
Here $K^2$ is the degree of the Del Pezzo (which is 9 minus the number of points blown up),
$\alpha, \ \beta$ and $\gamma$ are the numbers of
exceptional sheaves in each of the three blocks, and $x,\ y$ and
$z$ are the ranks of the sheaves in each
block (which one can prove to be equal within each block). This equation is invariant under ``block''
Seiberg dualities and is a special case for three-block collections of a Markov equation
for more general exceptional collections which is invariant under arbitrary Seiberg dualities \FengKK.
Equation \markov\ is also satisfied if one takes $x,\ y$ and $z$ to be the
ranks of the gauge groups within each block, provided the left hand side of the equation is
multiplied by the number $N$ of D3 branes we are describing.
These two possibilities are related by three ``block'' Seiberg
dualities \Karpov. We would like to mention that \markov\ has been interpreted as a consistency condition for
the vanishing of NSVZ beta-functions in \HerzogDJ.

The $n$th Del Pezzo has a discrete group of global diffeomorphisms
isomorphic to the exceptional group $E_n$. It is known that the induced action
of $E_n$ on  the quiver diagram  yields a symmetry of the diagram \Karpov. It might be
interesting to understand what kind of constraints it puts on the superpotential.

In the following sections we will display our quiver diagrams in
``block'' notation in order to avoid cluttering the pictures with
arrows.
When nodes are grouped together into a single block, there are no
arrows between them. Also, an arrow from one group to another group
signifies an arrow from each node in the first group to each node in
the second group. In the remainder we will also leave out the overall
trace of the superpotentials, it being understood that this has to be
added back to get a gauge invariant expression.

\subsec{Del Pezzo 3}

In this section we use the approach based on exceptional collections
to construct a quiver gauge theory for a local DP3. We check the answer we get
by Seiberg dualising to a known quiver gauge theory that was
constructed by toric methods.

We choose to use the exceptional collection that was found in
\Karpov. It is a particularly nice collection because it consists only
of line bundles and because it gives rise to
a superpotential with only cubic couplings.
To introduce it let us first fix some notation. DP3 can be
obtained as a blow-up of ${\bf P}^2$ at three generic points. The divisor
classes are then linear combinations of the three exceptional curves
$E_1,E_2$ and $E_3$, and the hyperplane class $H$, which satisfy the
following relations:
\eqn\aa{ E_i^2 = -1, \quad  H^2 = 1, \quad E_i \cdot E_j = 0, \quad
H \cdot E_i =
0.}
Then the exceptional collection of interest is given by the following
set of line bundles, presented in ``three-block'' form:
\eqn\dpthreeexc{ \matrix{
1.\ \cO && 2.\ \cO(H)                 && 4.\ \cO(2H - E_1 - E_2)  \cr
 && 3.\ \cO(2H - E_1 - E_2 - E_3) && 5.\ \cO(2H - E_1 - E_3)  \cr
  &&                          && 6.\ \cO(2H - E_2 - E_3)   }}
The point of this grouping of the sheaves is that there are no arrows
between any two members of the same block, and the number of arrows from
any member of a block to another block is the same. Now
to find the quiver diagram we proceed as outlined in
section 2. For each pair of nodes $i$ and $j$ we need to find  the dimensions
of ${\rm Ext}^k(E_i,E_j)$. In the case at hand  the $E_i$ are line
bundles and therefore the Ext's just reduce to the sheaf
cohomology groups $H^k(E_j \otimes E_i^*)$. The dimensions of these groups are
easily computed to be zero unless $k=0$.
Let us denote an arbitrary sheaf in the $i$th block by $E_{(i)}$. Then
the  cohomologies are given by:
\eqn\aa{
{\rm dim}\ \ext^0(E_{(1)},E_{(2)}) = 3, \quad
{\rm dim}\ \ext^0(E_{(2)},E_{(3)}) = 1, \quad
{\rm dim}\ \ext^0(E_{(1)},E_{(3)}) = 4.
}
To find the ranks of the
gauge groups, $n_i$, we need to satisfy the equation
\eqn\aa{ \sum_i n_i\, {\rm ch}(E_i) = {\rm ch}(\cO_p)}
where $\cO_p$ is the skyscraper sheaf over a point $p$ on the Del
Pezzo. This is
just the statement that the brane configuration has the correct charges of a
set of D3 branes filling the dimensions transverse to the Calabi-Yau,
but it
will also guarantee that we end up with an anomaly free theory
\CachazoSG. In the present case we can take the $n_i$ to be
\eqn\aa{ \{ n_1,n_2,n_3,n_4,n_5,n_6\} = \{ 1,-2,-2,1,1,1 \} .}
The
effect of the ``-2' on the quiver
diagram is that the arrows involving nodes 2 and 3 need to be reversed
(because we turned a brane into an anti-brane)
and the gauge groups at these nodes will be $U(2N)$ instead of $U(N)$.
The resulting quiver diagram is displayed below (see the
introduction to this section for an explanation of the block notation):
\bigskip
\centerline{\epsfxsize=0.72\hsize\epsfbox{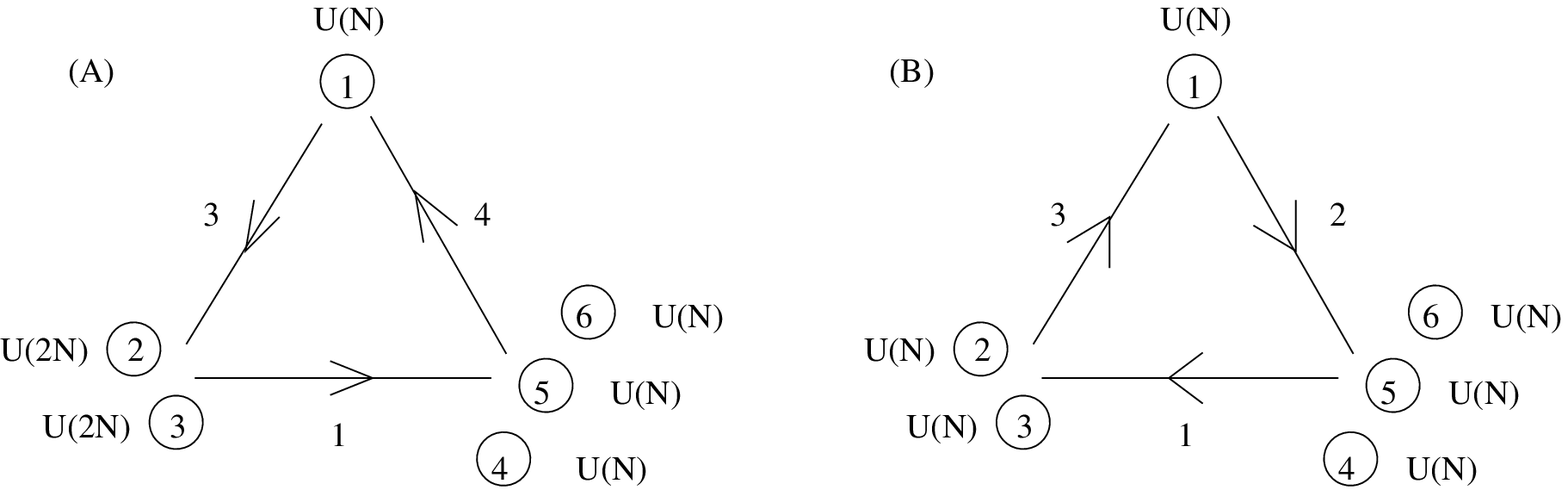}}
\noindent{\ninepoint\sl \baselineskip=8pt \fig\dpthree{}:
{\sl (A) Quiver for the exceptional collection in \dpthreeexc. (B) Seiberg dual theory,
also known as "model IV".}}
\bigskip


In order to compute the superpotential it is not sufficient to know the
dimensions of $H^0(E_j\otimes E_i^*)$; we also need to know the
generators. These can be represented as polynomials on
the underlying ${\bf P}^2$.

Let us write the coordinates on ${\bf P}^2$ as $[x_0,x_1,x_2]$ and fix the
points on ${\bf P}^2$ obtained by blowing down the exceptional curves as

\eqn\aa{ E_1 \sim [1,0,0], \quad E_2 \sim [0,1,0], \quad E_3 \sim
[0,0,1] .}
The generators we use are listed below:
\eqn\aa{ \eqalign{
X_{12} &= A_{12} \, x_{0} + B_{12} \, x_{1} + C_{12} \, x_{2} \cr
X_{13} &= A_{13} \, x_{1} x_{2} + B_{13}\,  x_{0} x_{2} + C_{13}\,  x_{0} x_{1} \cr
X_{24} &= x_{2} \cr
X_{25} &= x_{1} \cr
X_{26} &= x_{0} \cr
X_{34} &= 1 \cr
X_{35} &= 1 \cr
X_{36} &= 1 \cr
X_{14} &= A_{14} \, x_{1} x_{2} + B_{14}\, x_{0} x_{2} + C_{14}\,
x_{0} x_{1} + D_{14}\,  x_{2}^2 \cr
X_{15} &= A_{15} \, x_{1} x_{2} + B_{15}\, x_{0} x_{2} + C_{15}\,
x_{0} x_{1} + D_{15}\,  x_{1}^2 \cr
X_{16} &= A_{16} \, x_{1} x_{2} + B_{16}\, x_{0} x_{2} + C_{16}\,
x_{0} x_{1} + D_{16}\,  x_{0}^2 }}
Note that for $ X_{34} \ldots  X_{36}$ we have chosen generators that do
not seem to vanish in the right places. However these will pull back
to the right sections on Del Pezzo 3, and it is only the $x_i$
dependence that matters for our calculations. Now we can compute any three point
coupling we want by composing the generators around a loop. This gives
rise to the following superpotential:
\eqn\aa{ \eqalign{ W&= A_{21} X_{52} C_{15} + B_{12} X_{26} C_{61} +
B_{12} X_{24} A_{41} + C_{12} X_{25} A_{51} \cr &+ A_{12} X_{24}
B_{41} +C_{12} X_{26} B_{61} + A_{12} X_{26} D_{61} + B_{12} X_{25} D_{51}
\cr &+ C_{12} X_{24} D_{41}+A_{13} X_{34} A_{41}+A_{13} X_{35} A51+A_{13} X_{36}
A_{61} \cr &+ B_{13} X_{34} B_{41} + B_{13} X_{35} B_{51} +B_{13} X_{36}
B_{61} \cr &+C_{13} X_{34} C_{41} + C_{13} X_{35} C_{51} + C_{13} X_{36}
C_{61}  .}}
This superpotential can be compared with existing results in the
literature.
After applying Seiberg dualities on nodes 2 and 3 we get the quiver
diagram depicted in \dpthree B. It is known as model IV in
\refs{\BeasleyZP,\FengBN}. The
superpotential obtained
through Seiberg duality is
\eqn\knowndpsuper{ \eqalign{
W_{\rm dual} =
&-A_{12} A_{41} X_{24}-A_{12} A_{51} X_{25}-B_{12} B_{41} X_{24}-A_{61}
B_{12} X_{26}\cr
&-B_{51} C_{12} X_{25}-B_{61} C_{12} X_{26}+A_{13} B_{41}
X_{34}+A_{13} B_{51} X_{35}\cr
&+A_{41} B_{13} X_{34}+B_{13} B_{61}
X_{36}+A_{51} C_{13} X_{35}+A_{61} C_{13} X_{36}.}}

After a relabelling of the nodes and some simple field redefinitions
this agrees exactly with \refs{\BeasleyZP,\FengBN}.

\subsec{Del Pezzo 4}

In this subsection we analyse the first non-toric example. Its
superpotential will again be completely cubic. The
notation will be similar to the previous subsection except that we add
and extra exceptional curve $E_4$ which blows down to
\eqn\aa{ E_4 \sim [1,1,1] .}
Now we proceed with the construction of the quiver.
The exceptional collection from \Karpov\ is given by:
\eqn\dpfourexc{ \matrix{
1.\ \cO && 2.\ F && 3.\ \cO(H) \cr
  &&           && 4.\ \cO(2H - E_2 - E_3 - E_4) \cr
   &&          && 5.\ \cO(2H - E_1 - E_3 - E_4) \cr
   &&          && 6.\ \cO(2H - E_1 - E_2 - E_4) \cr
   &&          && 7.\ \cO(2H - E_1 - E_2 - E_3) \cr  }}
Here $F$ is a rank two bundle defined as the unique extension of the
line bundles $\cO(H)$ and $\cO(2H - \sum_i E_i)$:
\eqn\aa{
0 \to \cO(2H - \sum_i E_i) \to F \to \cO(H) \to 0 .}
The cohomologies for this collection are:
\eqn\aa{
{\rm dim}\ \ext^0(E_{(1)},E_{(2)}) = 5, \quad
{\rm dim}\ \ext^0(E_{(2)},E_{(3)}) = 1, \quad
{\rm dim}\ \ext^0(E_{(1)},E_{(3)}) = 3.
}
Because of the presence of a rank two bundle our computations
differ from those in the previous subsection. All computations for
this case were performed in by embedding the Del Pezzo in projective
space using its anti-canonical embedding and constructing the sheaves on it
in Macaulay2 \Macaulay. Let us make some comments about how this can
be done, since it may be useful for other computations.
To get the equations of the Del Pezzo one picks a linearly independent set of cubic
polynomials on ${\bf P}^2 $ which vanish on the points we want to blow
up. There are six of these and we can denote them by $c_1 , \ldots,
c_6$. Then through the map
\eqn\antican{ [x_{0},x_{1},x_{2}] \to [c_1(x_{0},x_{1},x_{2}), \ldots
,
c_6(x_{0},x_{1},x_{2})] }
our  ${\bf P}^2 $ parametrises a submanifold of  ${\bf P}^5 $. This map
seems not so well behaved near the marked points on  ${\bf P}^2 $ which we
wanted to blow up, since all the cubics are zero there but the origin
is excised from ${\bf C}^6$ when we form  ${\bf P}^5 $. The idea is that if
we find the smallest variety which contains the submanifold
parametrised by  ${\bf P}^2 $ then this variety will have the marked
points blown up into 2-spheres, and so will be our Del Pezzo surface.
Given a set of cubics, we can find this variety explicitly by
elimination theory.

Once we have the equations for the variety, we need to construct
the sheaves. These can be represented as modules over the coordinate
ring of the variety. For example if we want the sheaf $\cO(-E)$ where
$E$ is an exceptional curve obtained from blowing up the point $p$,
we can use elimination theory to find the  ``image'' of $p$ under the map
\antican. This yields a set of linear equations forming an ideal, and
the associated module is the module for the sheaf $\cO(-E)$.

Finally we need to build the sheaf $F$ by extension, find all the Ext
generators and compute the Yoneda pairings. In order to do this we
used ideas from \Smith. The point of that paper is that sheaf
Ext  can be computed as module Ext if we first truncate the modules. A
bound for the truncation was given in \Smith. For our computations
we found that frequently no truncation was needed at all. One
may now find the sheaf $F$ by computing the extension of the
truncated modules using the ``push-out'' construction, and calculate the
Yoneda pairings by computing compositions of Ext's (which for us were
just $\ext^0$'s) using maps between the modules.

The quiver diagram for the collection \dpfourexc\ is
\bigskip
\bigskip
\centerline{\epsfxsize=0.4\hsize\epsfbox{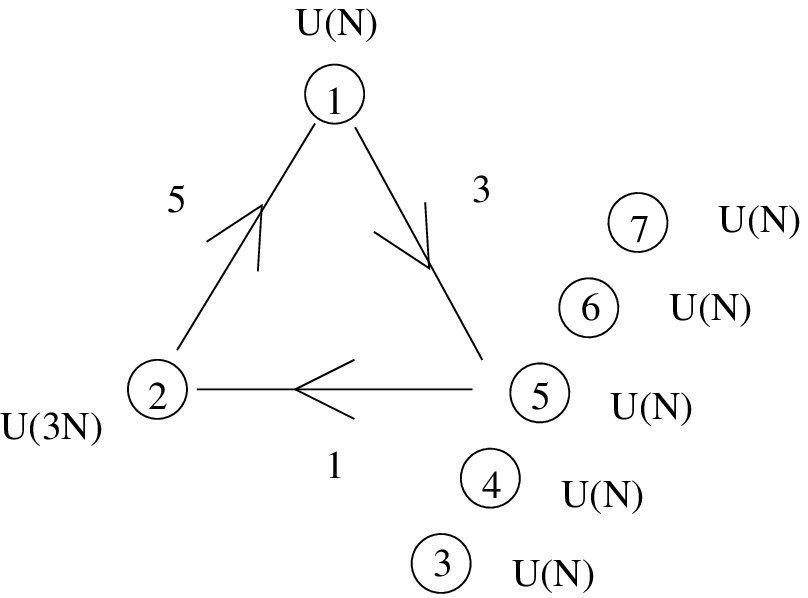}}
\noindent{\ninepoint\sl \baselineskip=8pt \fig\dpfour{}:
{\sl  Quiver diagram for the exceptional collection in \dpfourexc.}}
\bigskip
%
\noindent
The superpotential is found to be:
\eqn\aa{ \eqalign{ W =&
\ \
A_{13} X_{32} C_{21} + B_{13} X_{32} D_{21} + C_{31} X_{32} E_{21} + C_{14}
X_{42} A_{21} + B_{14} X_{42} B_{21} \cr
&+ B_{41} X_{42} C_{21} - A_{14}
X_{42} C_{21} + C_{41} X_{42} E_{21} + B_{14} X_{42} E_{21} - A_{14} X_{42}
E_{21} \cr
&+ C_{15} X_{52} A_{21} + A_{15} X_{52} B_{21} - C_{15} X_{52}
C_{21} - A_{15} X_{52} D_{21} - C_{15} X_{52} E_{21} \cr
&- B_{15} X_{52}
E_{21} + C_{16} X_{62} A_{21} + B_{16} X_{62} B_{21} + A_{16} X_{62} B_{21} -
A_{16} X_{62} C_{21} \cr
&- A_{16} X_{62} D_{21} - A_{16} X_{62} E_{21} +
C_{17} X_{72} A_{21} + B_{17} X_{72} B_{21} + A_{17} X_{72} B_{21} \cr
&-
A_{17} X_{72} D_{21} + B_{17} X_{72} E_{21} .}}
We would like to perform two checks on this result. First we would like
to show that the moduli space of this quiver is indeed Del Pezzo 4,
after that we will see how one can Higgs this theory down to Del Pezzo
3.

To find the moduli space we should write down the generators of the
ring of gauge invariant
operators and then impose the algebraic relations between them as well
as the constraints from the superpotential. We will actually only do
this for the case of a single D3-brane; the more general case can be
handled by thinking of the generators below as $N \times N$ matrices.
Our quiver diagram for Del
Pezzo 4 counts 75 generators, and from the superpotential we get 69
linear relations between them. We will take the six linearly
independent ones to be
\eqn\aa{\eqalign{
&   m_{67} = A_{17}\,  X_{72}\,  C_{21}, \quad
    m_{68} = B_{17}\,  X_{72}\,  C_{21}, \quad
    m_{69} = C_{17}\,  X_{72}\,  C_{21}, \cr
&       m_{73} = A_{17}\,  X_{72}\,  E_{21}, \quad
    m_{74} = B_{17}\,  X_{72}\,  E_{21}, \quad
    m_{75} = C_{17}\,  X_{72}\,  E_{21}\, .}}
Now there are 3 'obvious' quadratic relations between them and 2
non-obvious ones that make use of the F-term equations:
\eqn\aa{ \eqalign{
0 &=
{{m}}_{{69}} {{m}}_{{74}}-{{m}}_{{68}} {{m}}_{{75}}, \cr
0 &=
{{m}}_{{69}} {{m}}_{{73}}-{{m}}_{{67}} {{m}}_{{75}},\cr
0 &=
{{m}}_{{68}} {{m}}_{{73}}-{{m}}_{{67}} {{m}}_{{74}}, \cr
0 &=
{{m}}_{{73}} {{m}}_{{74}}+{{m}}_{{74}}^{2}-{{m}}_{{67}}
{{m}}_{{75}}-{{m}}_{{73}} {{m}}_{{75}}-{{m}}_{{74}} {{m}}_{{75}}, \cr
0 &=
{{m}}_{{67}} {{m}}_{{69}}-{{m}}_{{67}} {{m}}_{{74}}-{{m}}_{{68}}
{{m}}_{{74}}+ {{m}}_{{67}}
{{m}}_{{75}}+{{m}}_{{68}} {{m}}_{{75}} . }}
%
%
These give rise to a degree five surface in ${\bf P^5}$ which is our
Del Pezzo 4.

\bigskip
\centerline{\epsfxsize=1.0\hsize\epsfbox{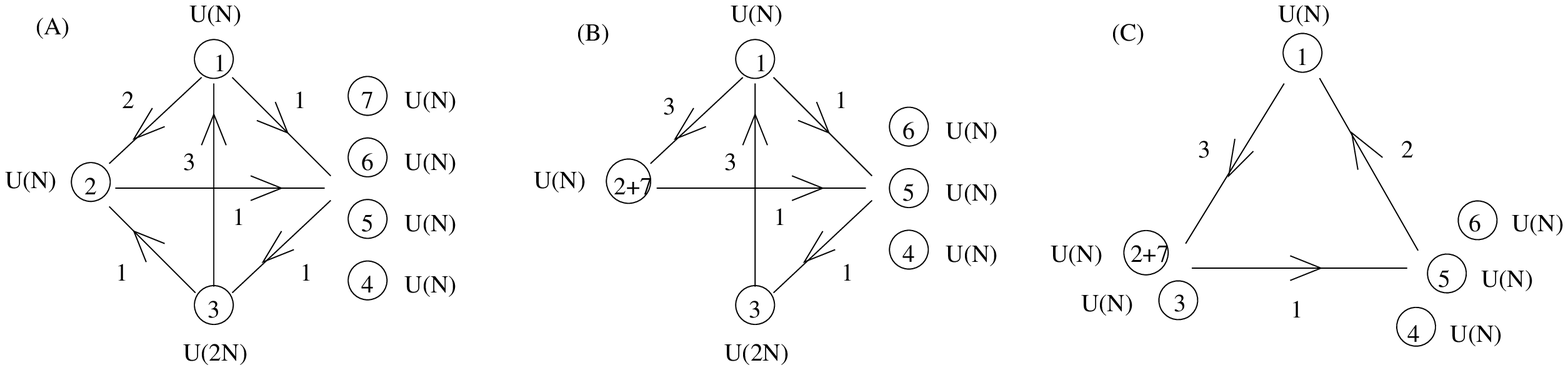}}
\noindent{\ninepoint\sl \baselineskip=8pt \fig\higgseddpfour{}
{\sl (A): DP4 quiver, dualised version of \dpfour. (B): condensing $X_{27}$
leads to a DP3 quiver. (C): by performing an additional Seiberg duality we arrive at "model IV"
for DP3.}}
\bigskip

The other check we would like to perform is Higgsing down the
theory in an appropriate way. The idea is make  a partial resolution of
the Calabi-Yau singularity such that the new singularity is the affine
cone over a Del Pezzo with fewer exceptional curves, and to compare
with existing result for the simpler case. In order to do this we have to write
down a baryonic operator whose VEV higgses the theory to a new
quiver where the  $E_4$ charges in the Chern characters have all disappeared.
It should be possible to do that directly
in the quiver we have considered, but this is likely to be somewhat complicated
because the $E_4$ charge is spread over several nodes.
Recall that equation \monodromy\ tells us what happens with the
Chern characters after a Seiberg duality.
So we will apply some Seiberg dualities until the $E_4$ charge in the
Chern characters appears in only two nodes linked by a single arrow
(so as
to avoid creating adjoints fields in the process of Higgsing
down).

The Chern characters of the original collection \dpfourexc\ are
given by
\eqn\dpfourch{ \matrix{
1.\ (1,0,0) && 2.\ -(2,3 H - \sum_i E_i,{1/ 2}) && 3.\
  (1,H,{1/ 2}) \cr
  &&           && 4.\ (1,2H - E_2 - E_3 - E_4,{1/2}) \cr
   &&          && 5.\ (1,2H - E_1 - E_3 - E_4,{1/ 2}) \cr
   &&          && 6.\ (1,2H - E_1 - E_2 - E_4,{1/ 2}) \cr
   &&          && 7.\ (1,2H - E_1 - E_2 - E_3,{1/ 2}) \cr  }}
The entries in each vector in the above indicate $({\rm rank},c_1,{\rm
ch}_2)$
of the sheaves. We also use an extra minus sign at node 2 so that the
sum of the Chern characters (with proper multiplicities)
is $(0,0,1)$, the Chern character for the
D3-brane.
Now we
can  achieve our objective by performing Seiberg dualities on
nodes 3 and 2, after which the quiver diagram takes the form in
figure \higgseddpfour A with superpotential
\eqn\aa{\eqalign { W = &
\ \ {A_{31}}\,{X_{14}}\,{X_{43}} + {C_{31}}\,{X_{14}}\,{X_{43}} +
{A_{12}}\,{A_{31}}\,{X_{24}}\,{X_{43}} +
{A_{12}}\,{C_{31}}\,{X_{24}}\,{X_{43}} \cr
&+
{B_{12}}\,{C_{31}}\,{X_{24}}\,{X_{43}} +
{X_{24}}\,{X_{32}}\,{X_{43}}
+   {C_{31}}\,{X_{15}}\,{X_{53}} -
{A_{31}}\,{B_{12}}\,{X_{25}}\,{X_{53}} \cr
&-
{A_{12}}\,{B_{31}}\,{X_{25}}\,{X_{53}} -
{B_{12}}\,{C_{31}}\,{X_{25}}\,{X_{53}} +
{X_{25}}\,{X_{32}}\,{X_{53}}
-   {1\over 2}{A_{31}}\,{X_{16}}\,{X_{63}} \cr
&-
{1\over 2}{B_{31}}\,{X_{16}}\,{X_{63}} -
{1\over 2}{C_{31}}\,{X_{16}}\,{X_{63}}-
{1\over 2}{A_{12}}\,{A_{31}}\,{X_{26}}\,{X_{63}}-
{1\over 2}{A_{12}}\,{B_{31}}\,{X_{26}}\,{X_{63}} \cr
&-
{1\over 2}{A_{12}}\,{C_{31}}\,{X_{26}}\,{X_{63}} +
{X_{26}}\,{X_{32}}\,{X_{63}} -
{1\over 2}{B_{31}}\,{X_{17}}\,{X_{73}}
-   {1\over 2}{C_{31}}\,{X_{17}}\,{X_{73}} \cr
&-   {1\over 2}{A_{12}}\,{B_{31}}\,{X_{27}}\,{X_{73}} +
{1\over 2}{A_{12}}\,{C_{31}}\,{X_{27}}\,{X_{73}} +
{X_{27}}\,{X_{32}}\,{X_{73}} }
}
If we do the Seiberg dualities on nodes 3 and 2 by applying mutations
to nodes $\{2\}$ and $\{4,5,6,7\}$ respectively, then the Chern
characters for the new quiver diagram take the following form:
\eqn\dpfoursdch{ \matrix{
1.\ (1,0,0) && 2.\ (1,2 H - \sum_i E_i,0) && 3.\
 - (1,H,{1/ 2})   && 4.\ (0,E_1,{1/ 2}) \cr
   &&     &&     && 5.\ (0,E_2,{1/ 2}) \cr
   &&     &&     && 6.\ (0,E_3,{1/ 2}) \cr
   &&     &&     && 7.\ (0,E_4,{1/ 2}) \cr  }}
The $E_4$ charge now only appears in nodes 2 and 7, and checking the
relative Euler characteristic shows that the link between them is
given by an $\ext^1$.
Now we may Higgs down to Del Pezzo 3 by turning on an expectation
value for $X_{27}$, which gives rise to a new exceptional collection
of sheaves:\foot{
The unique extension can be found by tensoring the fundamental
sequence
$0\to \cO(-E_4) \to \cO \to \cO_{E_4} \to 0$ with $\cO(2H - E_1 - E_2
- E_3)$, which is consistent with the Chern characters.}
\eqn\dpfourhiggsch{ \matrix{
1.\ (1,0,0) && 2+7.\ (1,2 H - E_1 - E_2 - E_3,{1/ 2}) && 3.\
  -(1,H,{1/ 2})   && 4.\ (0,E_1,{1/ 2}) \cr
   &&     &&     && 5.\ (0,E_2,{1/ 2}) \cr
   &&     &&     && 6.\ (0,E_3,{1/ 2}) \cr  }}
The resulting quiver is drawn in \higgseddpfour B.

We can
do a final Seiberg duality on node 3 after which we arrive at the
quiver diagram of model IV. If we perform a mutation on nodes
$\{4,5,6\}$ then we get the Chern characters for the line bundles
\eqn\dpfourhiggsexc{ \matrix{
1.\ \cO && 2+7.\ \cO(2 H - E_1 - E_2 - E_3,) && 4.\ \cO(H-E_1) \cr
   &&    3.\ \cO(H)    && 5.\ \cO(H-E_2) \cr
   &&                  && 6.\ \cO(H-E_3) \cr  }}
Upon some field redefinitions the superpotential reduces
exactly to the expected known superpotential for model IV, given
eg. in equation \knowndpsuper\  above. Moreover a further Seiberg
duality on nodes $2+7$ and $3$ gives rise to the exceptional
collection \dpthreeexc\ we used in the previous subsection.

\subsec{Del Pezzo 5}

For Del Pezzo surfaces of with more than four exceptional curves
we have to deal with a new
phenomenon: these surfaces have a complex structure moduli
space. While our computations are insensitive to the K\"ahler
structure, they definitely do depend on the complex structure so we
expect these moduli to make an appearance in the superpotential.

The appearance of complex structure moduli is easy to understand
in the description of Del Pezzos we have used, as blow-ups of
${ \bf P}^2$. The set of coordinate transformations preserving the
complex structure of ${ \bf P}^2$ is the group $PGl(3,C)$, which has 8
complex parameters. The complex structure of the Del Pezzos is completely
determined by specifying which points on ${ \bf P}^2$ get blown up.  To
specify a point on ${ \bf P}^2$ one needs two complex parameters.
Therefore, the first four marked points can always be fixed at some
chosen reference points, but then we have
used up the coordinate transformations and for each additional
point we have a choice of two complex parameters each of which
gives rise to a different complex structure.

With this in mind, for DP5 we add a fifth exceptional curve $E_5$
which blows down to a point with floating coordinates. For
convenience we use projective coordinates to parametrise this marked point
and therefore the complex structure moduli space, $E_5 \sim [z_0,z_1,z_2]$.

The exceptional collection of \Karpov\ is given by the following collection of
line bundles:
\eqn\dpfiveexc{ \matrix{
1.\ \cO(E_5) && 3.\ \cO(H) && 5.\ \cO(2H - E_1 - E_2) \cr
2.\ \cO(E_4) && 4.\ \cO(2H - E_1 - E_2 - E_3) && 6.\ \cO(2H - E_2 - E_3) \cr
 &&  && 7.\ \cO(2H - E_1 - E_3) \cr
 &&  && 8.\ \cO(3H - \sum_{i=1}^5 E_i)  }}
The cohomologies for this collection are:
\eqn\aa{
{\rm dim}\ \ext^0(E_{(1)},E_{(2)}) = 2, \quad
{\rm dim}\ \ext^0(E_{(2)},E_{(3)}) = 1, \quad
{\rm dim}\ \ext^0(E_{(1)},E_{(3)}) = 3.
}
The quiver diagram for this collection is displayed below:
\bigskip
\centerline{\epsfxsize=0.4\hsize\epsfbox{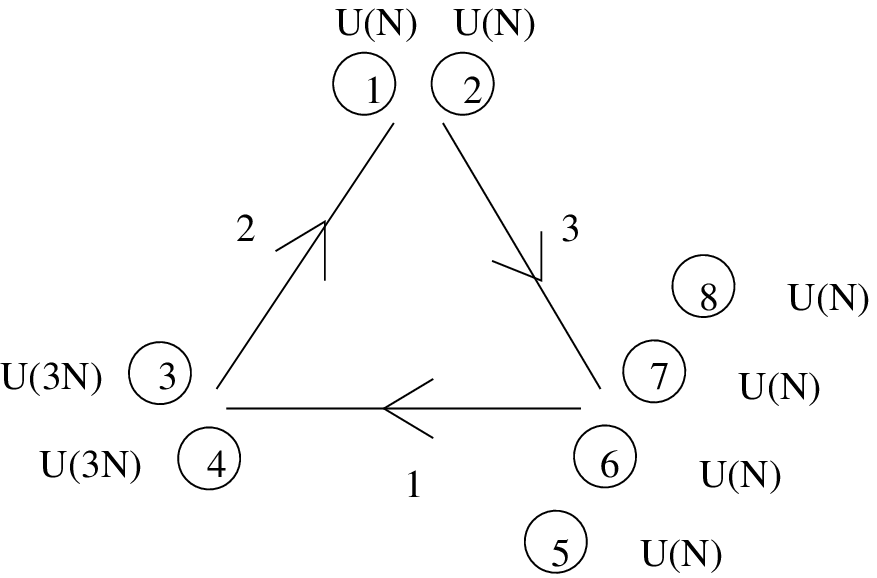}}
\noindent{\ninepoint\sl \baselineskip=8pt \fig\dpfive{}:
{\sl Quiver diagram corresponding to the exceptional collection in \dpfiveexc. }}
\bigskip%
This quiver diagram allows for cubic terms and sextic terms in the
action. The calculation for the cubic terms is very similar to the
computation for DP3, so we will not
write down the details here. The answer we get is
\eqn\aa{ \eqalign{ W_{\rm cubic}=& \ \ X_{53} A_{31} (-
{z_{0}\over z_{2}} A_{15} + {z_{1}\over z_{2}} B_{15}) + X_{53}
B_{31} B_{15} + X_{63} A_{31} (-{z_{0}\over z{2}} B_{16} +
{z_{0}\over z_{2}} C_{16} ) \cr &+ X_{63} B_{31} (- {z_{0}\over z_{1}}
A_{16}
+ {z_{0}\over z_{1}} C_{16} ) + X_{73} A_{31} (-{z_{0}\over z_{2}} B_{17} + {z_{1}\over z_{2}}
C_{17} ) \cr &+ X_{73} B_{31} C_{17} + X_{83} A_{31} ({z_{0}\over z_{2}} A_{18} + {z_{0}
z_{1}
\over z_{2} (z_{1}-z_{2})} B_{18} + {z_{1} (z_{0} - z_{2}) \over z_{0} (z_{1} - z_{2})} C_{18} )
\cr &+ X_{83} B_{31} (-{z_{0}\over z_{2} - z_{1}} B_{18} + {z_{2} (z_{0} - z_{2}) \over
z_{0}
(z_{1} - z_{2}) } C_{18} ) + X_{53} A_{32} (- A_{25} + B_{25}) \cr &+ X_{53} B_{32} B_{25} +
X_{63}
A_{32} (-B_{26} + C_{26}) + X_{63} B_{32} (-A_{26} + C_{26}) \cr & + X_{73} A_{32} (-B_{27} +
C_{27}) + X_{73} B_{32} C_{27} \cr & + X_{83} A_{32} ({z_{1} (z_{2} - z_{0})\over z_{0} (z_{1} -
z_{2})} A_{28} + B_{28} + {z_{0} \over z_{2} - z_{1}} C_{28} ) \cr &+ X_{83} B_{32} ({z_{1}
(z_{2}
- z_{0}) \over z_{0} (z_{1} - z_{2})} A_{28} + {z_{0}\over z_{2} - z_{1}} C_{28}) \cr &+
X_{54}
A_{41} (-{z_{0}\over z_{2}} A_{15} + {z_{1} \over z_{2}} B_{15}) + X_{54} B_{41} (-{z_{0}\over
z_{2}} A_{15} + {z_{0} \over z_{2}} C_{15}) + X_{64} A_{41} A_{16} \cr &+ X_{64} B_{41} B_{16} +
X_{74} A_{41} A_{17} + X_{74} B_{14} C_{17} - X_{84} A_{41} C_{18} \cr &+ X_{54} A_{42}( - A_{25} +
B_{25}) + X_{54} B_{24}(-A_{25} + C_{25}) + X_{64} A_{42} A_{26} \cr
&+ X_{64} B_{42} B_{26}
+X_{74} A_{42} A_{27}
+ X_{74} B_{42} C_{27} + X_{84} A_{42} A_{28} \cr &+ X_{84} B_{41}( {z_{0} ( z_{2} - z_{0})
\over z_{2} (z_{1} - z_{0})} A_{18} + {z_{0}^2 \over z_{2} (z_{0} - z_{1})} B_{18}) \cr &+
X_{84} B_{42} ( {z_{0 }(z_{2} - z_{0})\over z_{2} (z_{1} - z_{0})} B_{28} + {z_{0}^2 \over
z_{2}
(z_{1} - z_{0})} C_{28} ) . \cr}}
The sextic terms a priori may present a problem. Since there are
no $\ext^1$'s among the cohomologies and we do not have a
prescription for computing higher order couplings without
$\ext^1$'s, we cannot find the coefficients of the allowed sextic
terms from first principles. On the other hand, since all cubic terms appear in
the superpotential and therefore have R-charge equal to 2 and dimension 3, any sextic term would
have dimension 6 and so its coefficient in the superpotential would be
dimensionful. Such a term simply can't be present since we're considering the far IR
physics which is given by an (interacting) scale-invariant theory. One may also take a
limit in which the generic Del Pezzo becomes a toric surface and
compare with toric computations which can be done independently.
In this limit one finds only cubic terms. So we conclude that the
cubic terms provide the full answer.

Let us do some checks on the answer we have obtained.
The first test of the superpotential is a computation of the moduli
space. To reduce the number of gauge invariant operators we perform
Seiberg dualities on nodes 3,4,5 and 6 and make some field redefinitions
after which we arrive at the
quiver in \octa A with purely quartic couplings:
\eqn\aa{\eqalign{W = &
\ \ X_{27} X_{73} X_{35} X_{52}  + X_{27} X_{73} X_{36} X_{62}
+ X_{27} X_{74} X_{45} X_{52}  + X_{27} X_{74}
X_{46} X_{62}  \cr
&+ \! X_{17} X_{73} X_{36} X_{61}  + \! z_{1} X_{17} X_{74} X_{45} X_{51}
+\! z_{1} X_{17} X_{74} X_{46} X_{61}
+\! z_{2} X_{17} X_{73} X_{35} X_{51} \cr
& - (z_{1} - 1) X_{18} X_{83} X_{36} X_{61}
 + z_{1} (z_{1} - 1) X_{18} X_{84} X_{46} X_{61} \cr
& \qquad  + z_{1} (z_{1} - z_{2}) X_{18} X_{84} X_{45} X_{51}
+ (z_{2 }- z_{1}) X_{18} X_{83} X_{35} X_{51} \cr
& +
  (z_{1} - 1) X_{28} X_{83} X_{36} X_{62}
+ (z_{1} - z_{2}) X_{28} X_{83} X_{35} X_{52} \cr
& \qquad - z_{2} (z_{1} - 1) X_{28} X_{84} X_{46} X_{62}
+ (z_{2} - z_{1}) X_{28} X_{84} X_{45} X_{52}  .}}
The coordinate $z_{0}$ has been set to one in this expression, but may
of course always be restored by replacing $z_{1} \to z_{1}/z_{0}$ and
$z_{2} \to z_{2}/z_{0}$. When we compute the moduli space of this theory we seem to find two
components. That is, the ideal $I$ of relations we find can be
expressed as the intersection $I = I_1 \cap I_2$ where $I_1$
contains the equations for the affine cone over the
Del Pezzo and $I_2$ is another ideal
which geometrically contains only one point, the singular point of the
Calabi-Yau.\foot{
If the Calabi-Yau develops non-isolated singularities one finds
additional branches. Their interpretation is that the D3 branes can
split into fractional branes which can move along the singularities.}
Geometrically therefore the moduli space is exactly as expected.
%
%
%
For the $I_1$
component all the gauge invariant operators can be expressed in terms
of
\eqn\aa{\eqalign{
& m_{11} = X_{18} X_{83} X_{36} X_{61},
\quad m_{14} = X_{28} X_{84} X_{45} X_{52}, \quad
m_{16} = X_{28} X_{84} X_{46} X_{62}, \cr
& m_{12} = X_{28} X_{83} X_{36} X_{62}, \quad m_{15} = X_{18} X_{84} X_{46} X_{61} .}}
There is one obvious relation between these and one non-obvious
relation:
\eqn\dpfiveeqn{\eqalign{
0 &= m_{12} m_{15} - m_{11} m_{16}, \cr
0 &= z_{1} (z_{1} - z_{2}) m_{11} m_{14} - (z_{1} - z_{2}) m_{12}
m_{14}
- (z_{1} - z_{2})^2 m_{14}^2 -z_{1}(z_{1} - z_{2}) m_{14} m_{15} \cr
& \qquad
+ z_{1}(z_{1} - 1) m_{11}m_{16}
- z_{2}(z_{1} - 1) m_{12} m_{16} - 2 z_{2}(z_{1} - 1)(z_{1} -
z_{2})m_{14} m_{16} \cr
& \qquad
- z_{1} z_{2}(z_{1} - 1)m_{15} m_{16} + z_{2}^2(z_{1} - 1) m_{16}^2 .}
}
It would be interesting to check if this Del Pezzo has the expected
complex structure.
%
\bigskip
\centerline{\epsfxsize=1.00\hsize\epsfbox{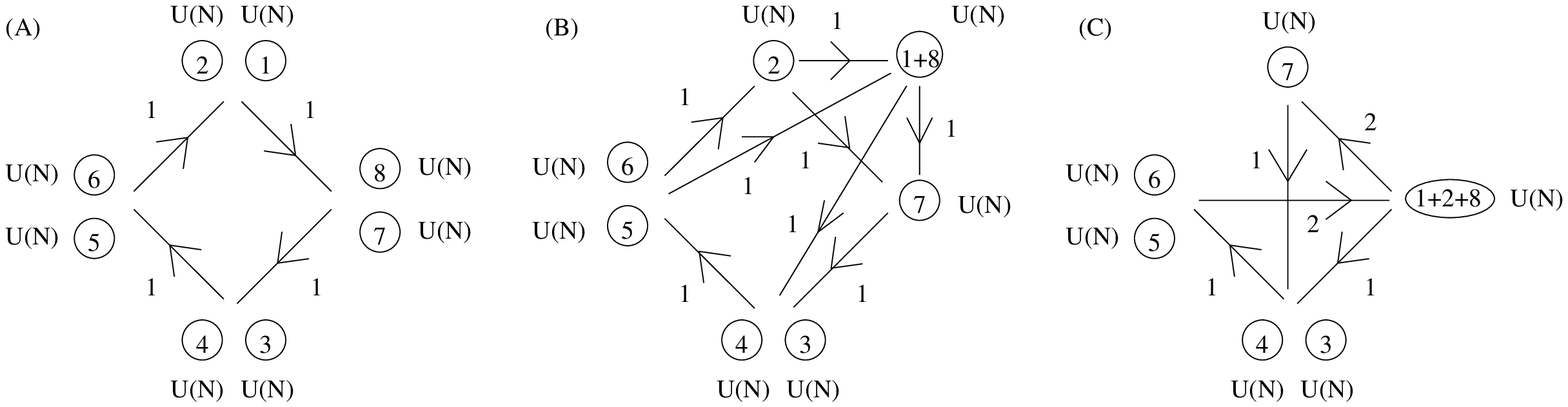}}
\noindent{\ninepoint\sl \baselineskip=8pt \fig\octa{}
{\sl (A): Quiver obtained by dualising \dpfive. (B): Giving a VEV to $X_{18}$
in (A), which results in a quiver for Del Pezzo 4. (C): Further
condensing $X_{28}$ gives rise to ``model III'' for Del Pezzo 3.}}
\bigskip%
Another possible way of testing the superpotentials is by
Higgsing down to known Del Pezzo's as we did for Del Pezzo 4. The
simplest way to do that in the present case is by giving an
expectation value to $X_{18}$ in the quiver of \octa A, after
which we get the quiver theory in \octa B\ without having to do any
integrating out. One can
straightforwardly check that the moduli space of this quiver is
exactly Del Pezzo 4. One may Higgs it down further to Del Pezzo 3 by
condensing $X_{28}$. This quiver is known as ``model III''
\refs{\BeasleyZP,\FengBN}, and the superpotential we get coincides with the known
superpotential after field redefinitions.

The quiver theory in \octa B is related to the Del Pezzo 4 quiver
 we found in
 the previous subsection. Namely we can start with \dpfour\ and apply
 Seiberg dualities on nodes 2,3,4,2,5.

Finally one can take a limit in which some of the marked points
approach each other. In such a limit the shrinking 4-cycle may cease to have
a positive curvature anti-canonical bundle and so would therefore
no longer be a Del Pezzo surface, but one may obtain cones over toric
surfaces in such limits or sometimes even orbifolds. For Del Pezzo
5 there is a limit in which the Calabi-Yau turns into the $Z_2
\times Z_2$ orbifold of the conifold, and another in which it becomes
a partial toric resolution of the orbifold singularity ${\bf C}^3/Z_3 \times
 Z_3$
 and hence we may compare
with known answers in the literature. We have not carried out the
details because the field rescalings are rather involved, but it
should be possible to check this.

\subsec{Del Pezzo 6}

For Del Pezzo 6  two three-block models were given
in \Karpov. We can compute the cubic couplings in either without
too much difficulty, but we will stick to the collection of line
bundles because it is easier to get the dependence on complex
structure moduli that way. We blow up a sixth point on ${\bf P^2}$
with floating coordinates
\eqn\aa{ E_6 \sim [w_{0},w_{1},w_{2}]. }
The exceptional sheaves of the first collection in \Karpov, which we
might call `` model 6.1'',  are:
\eqn\dpsixexc{\matrix{
1.\ \cO(E_4) && 4.\ \cO(H-E_1) && 7.\ \cO(2H - E_1 - E_2 - E_3) \cr
2.\ \cO(E_5) && 5.\ \cO(H-E_2) && 8.\ \cO(H) \cr
3.\ \cO(E_6) && 6.\ \cO(H-E_3) && 9.\ \cO(3H - \sum_{i=1}^6 E_i)  \cr
}}
with cohomologies given by
\eqn\aa{
{\rm dim}\ \ext^0(E_{(1)},E_{(2)}) = 1, \quad
{\rm dim}\ \ext^0(E_{(2)},E_{(3)}) = 1, \quad
{\rm dim}\ \ext^0(E_{(1)},E_{(3)}) = 2.
}
The quiver for this collection is displayed in \dpsix A.
\bigskip
\centerline{\epsfxsize=0.7\hsize\epsfbox{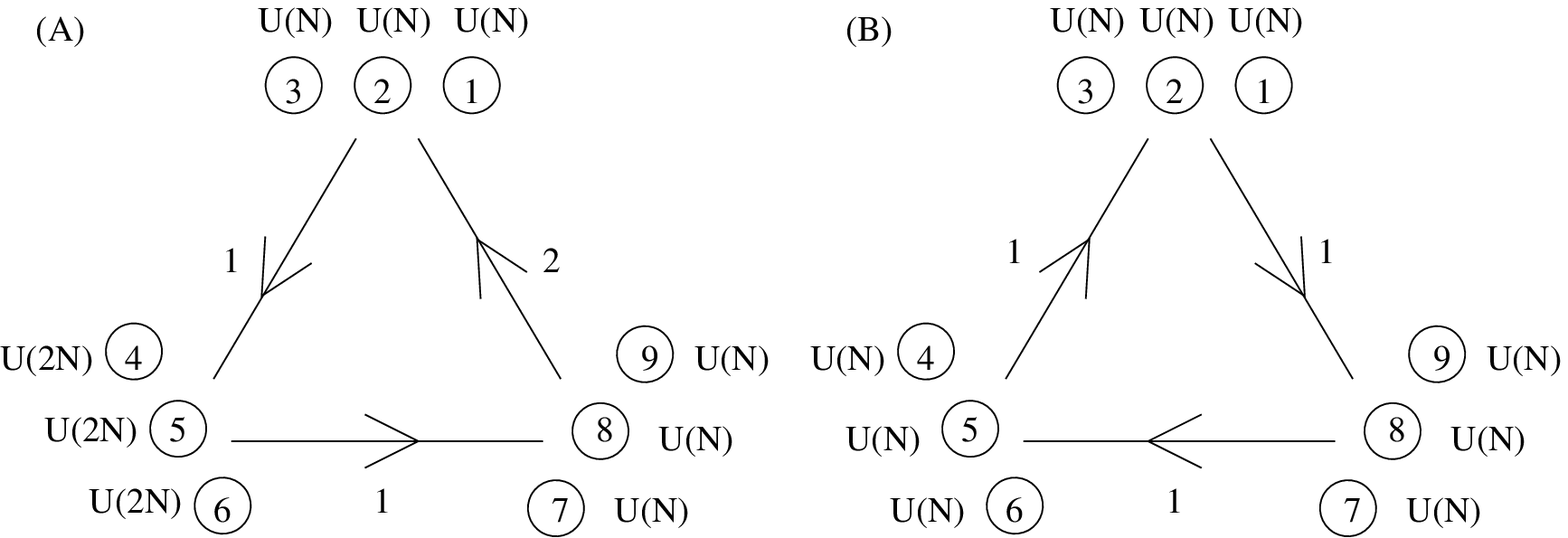}}
\noindent{\ninepoint\sl \baselineskip=8pt \fig\dpsix{}
{\sl (A): Quiver diagram corresponding to the exceptional collection in
 \dpsixexc.
(B): Seiberg dual quiver; also the quiver diagram for the orbifold
 ${\bf C^3}/{Z_3 \times Z_3}$. Notice the similarity between these two
 quivers and the ${\bf P}^2$ quivers in \Ptwo.}}
\bigskip%
Before writing down the superpotential let us introduce the following shorthand notation:
\eqn\aa{\eqalign{
f_{1} &= (z_{1} - z_{0}) z_{2} w_{0}^2 + (z_{0}^2 - z_{1} z_{2}) w_{0}
w_{2}
+ z_{0} (z_{2} - z_{0}) w_{1} w_{2} \cr
f_{2} &= z_{0} (z_{1} - z_{2}) w_{1} w_{2} + (z_{0} - z_{1}) z_{2}
w_{0} w_{1}
+ (z_{2} - z_{0}) z_{1} w_{0} w_{2} \cr
f_{3} &= -z_{1} (z_{2} - z_{1})w_{0} w_{2}
- (z_{0} - z_{1}) z_{2} w_{1}^2 - (z_{1}^2 - z_{0} z_{2}) w_{1} w_{2}
.}}
The cubic terms in the superpotential for model 6.1 are
\eqn\aa{\eqalign{
W &= \left( -{a_{71}} + {b_{71}} \right) \,{x_{14}}\,{x_{47}}  +
{x_{24}}\,{x_{47}}\, \left( {b_{72}} -
{{a_{72}}\,{z_{1}}\over{z_{2}}} \right)
+    \left(
{b_{73}} - {{a_{73}}\,{w_{1}}\over{w_{2}}} \right)
\,{x_{34}}\,{x_{47}}  \cr
&
+ {a_{81}}\,{x_{14}}\,{x_{48}}
+ {a_{82}}\,{x_{24}}\,{x_{48}}
+
{a_{83}}\,{x_{34}}\, {x_{48}}  \cr
&
+\left( {b_{91}}  +
{{a_{91}}\, {f_{1}}\over{f_{2}}}\right) \,{x_{14}}\, {x_{49}}
+
\left({b_{92}}      +
{{a_{92}}\,{f_{1}}\over{f_{2}}}  \right)  \,{x_{24}}\,{x_{49}}    +
\left( {b_{93}}         +
{{a_{93}}\,{f_{1}}\over{f_{2}}} \right) \,{x_{34}}\, {x_{49}}  \cr
&
+
{b_{71}}\,{x_{15}}\,{x_{57}} +    {b_{72}}\,{x_{25}}\,{x_{57}} +
{b_{73}}\,{x_{35}}\,{x_{57}} \cr
&
+    {b_{81}}\,{x_{15}}\,{x_{58}} +
{b_{82}}\,{x_{25}}\,{x_{58}} +    {b_{83}}\,{x_{35}}\,{x_{58}} \cr
&+
\left( {b_{91}}      +
{{a_{91}}\,{f_{2}}\over{f_{3}}} \right) \,{x_{15}}\,{x_{59}}
+
\left( {b_{92}}   +
{{a_{92}}\, {f_{2}}\over{f_{3}}}\right)\,{x_{25}}\, {x_{59}}
+
\left( {b_{93}}   +
{{a_{93}}\,{f_{2}}\over{f_{3}}} \right)\,{x_{35}}\,{x_{59}} \cr
&+
{a_{71}}\,{x_{16}}\,{x_{67}}  + {a_{72}}\,{x_{26}}\, {x_{67}}  + {a_{73}}
\,{x_{36}}\, {x_{67}}  \cr
&
+ \left( -{a_{81}} +  {b_{81}} \right)
\,{x_{16}}\,{x_{68}}   + {x_{26}}\,{x_{68}}\, \left( {b_{82}} -
{{a_{82}}\,{z_{0}}\over{z_{1}}} \right)+ \left( {b_{83}}  -
{{a_{83}}\,{w_{0}}\over{w_{1}}} \right) \,{x_{36}}\, {x_{68}}  \cr
&+
\left( -{a_{91}} +  {b_{91}} \right) \,{x_{16}}\,{x_{69}}  + \left(
{b_{93}} -   {{a_{93}}\,{w_{0}}\over{w_{1}}} \right) \,{x_{36}}\,
{x_{69}}    + {x_{26}}\,{x_{69}}\,
\left( {b_{92}} - {{a_{92}}\,{z_{0}}\over {z_{1}}} \right)
}}
As for Del Pezzo 5 we expect this to be the full answer.

In order to do the moduli space computation we go to a new quiver
by Seiberg dualities on nodes 4,5,6, after which we get \dpsix B
with superpotential
\eqn\aa{\eqalign{
W &= X_{17} (X_{41} X_{74}-X_{51} X_{75}+X_{61} X_{76}) \cr
&
+X_{18} (X_{41} X_{84}-X_{51} X_{85}+X_{61} X_{86}) \cr
&
+X_{19} ((-f_{2}^2 -f_{2} f_{3}) X_{41} X_{94}
+(f_{1} f_{3} +f_{2} f_{3} )X_{51} X_{95}
+(f_{2}^2 -f_{1} f_{3}) X_{61} X_{96}) \cr
&
+X_{27} (X_{62} X_{76} z_{1}
+X_{42} X_{74} z_{2}-X_{52} X_{75} z_{2}) \cr
&
+X_{28} (X_{42} X_{84} z_{0}-X_{52} X_{85} z_{1}+X_{62} X_{86} z_{1})
\cr
&
+X_{29} ((-f_{2} f_{3}  z_{0}-f_{2}^2  z_{1}) X_{42} X_{94}
+(f_{2} f_{3}  z_{0}+f_{1} f_{3}  z_{1}) X_{52} X_{95} ) \cr
& + X_{29}( z_{1}(f_{2}^2  -f_{1} f_{3} ) X_{62} X_{96})
 +X_{37} (w_{2} (X_{43} X_{74}-X_{53} X_{75})+w_{1} X_{63} X_{76})
\cr
&
+X_{38} (w_{0} X_{43} X_{84}-w_{1} X_{53} X_{85}+w_{1} X_{63} X_{86})
+X_{39} ((-f_{2} f_{3} w_{0}\! -f_{2}^2 w_{1}) X_{43} X_{94}) \cr
& + X_{39}(
(f_{2} f_{3} w_{0}\! +f_{1} f_{3} w_{1}) X_{53} X_{95}
+ w_{1}(f_{2}^2\!  -f_{1} f_{3} )X_{63} X_{96})
}}
When we calculate the ideal of relations we find, just as for Del
Pezzo 5, that it can be expressed as $I = I_1 \cap I_2$ where $I_1$
gives a cubic relation in four variables and $I_2$ only contains the
origin of moduli space geometrically. As before to get the locus of
the moduli space one should
take the radical of $I$, leaving us only with a cubic equation.
We were not able to get a general expression involving the complex
structure parameters as in \dpfiveeqn, but we can find the cubic
equation for any given complex structure. The 207 gauge invariant
operators can all be solved for in terms of
\eqn\aa{
p_{23} = X_{29} X_{95} X_{52} ,
\quad p_{24} = X_{39} X_{95} X_{53}  ,
\quad p_{26} = X_{29} X_{96} X_{62} ,
\quad p_{27} = X_{39} X_{96} X_{63}  .}
If for example the complex structure is given by
$[z_{0},z_{1},z_{2}] = [1,3,5]$ and $[w_{0},w_{1},w_{2}] =
[1,2,-2]$, we get
\eqn\aa{\eqalign{ 0 &= {{p}}_{{23}}\,
{{p}}_{{24}}\, {{p}}_{{26}}+{4\over 43}\, {{p}}_{{24}}^{2}\,
{{p}}_{{26}} +{105\over 731}\, {{p}}_{{24}}\,
{{p}}_{{26}}^{2}+{5\over 43}\, {{p}}_{{23}}^{2}\, {{p}}_{{27}}
+{20\over 43}\, {{p}}_{{23}}\, {{p}}_{{24}}\, {{p}}_{{27}} \cr
&-{825\over 731}\, {{p}}_{{23}}\, {{p}}_{{26}}\, {{p}}_{{27}}
-{10\over 731}\, {{p}}_{{24}}\, {{p}}_{{26}}\, {{p}}_{{27}}
-{350\over 731}\, {{p}}_{{23}}\, {{p}}_{{27}}^{2}. } }
We recognise the well-known realisation of Del Pezzo 6 as a degree 3 surface in ${\bf P}^3$.
%
\bigskip
\centerline{\epsfxsize=0.73\hsize\epsfbox{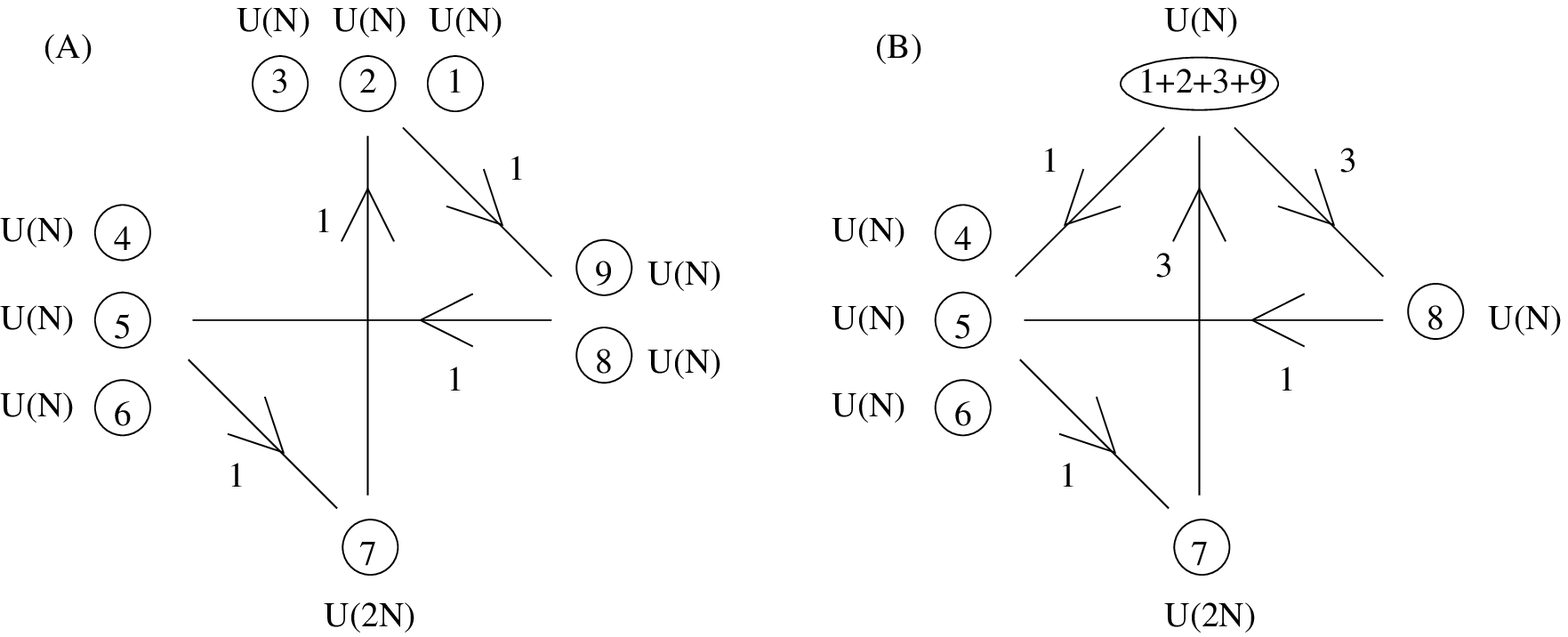}}
\noindent{\ninepoint\sl \baselineskip=8pt \fig\higgsdpsix{}
{\sl (A): Dual to \dpsix B by Seiberg duality on node 7,
 superpotential given in \higgsdpsixpot. Upon Higgsing
down we get quiver theories for DP5,DP4,DP3.
(B): DP3 quiver, related to model IV by duality on node 7.}}
\bigskip%

As usual we can also try to Higgs down to other Del Pezzo's. One seems
to get the simplest result by first Seiberg dualising on node 7 in
\dpsix B, after which one arrives at \higgsdpsix A with superpotential
\eqn\higgsdpsixpot{\eqalign{
W &= -X_{18}\, X_{47}\, X_{71}\, X_{84}
- {w_{0}\over w_{2}} \, X_{38}\, X_{47}\, X_{73}\, X_{84} -
      X_{18}\, X_{57}\, X_{71}\, X_{85} \cr
&
-\! {w_{1}\over w_{2}} \, X_{38}\, X_{57}\, X_{73}\, X_{85}
-\!
X_{18}\,X_{67}\,X_{71}\,X_{86}- \!X_{28}\,X_{67}\,X_{72}\,X_{86}
-\!X_{38}\,X_{67}\,X_{73}\,X_{86}\cr
&
+f_{2}^2\,X_{19}\,X_{47}\,X_{71}\,X_{94}
+f_{2}\,f_{3}\,X_{19}\,X_{47}\,X_{71}\,X_{94}
+{f_{2}\,f_{3}\,w_{0}\over w_{2}}\,X_{39}\,X_{47}\,X_{73}\,X_{94} \cr
&+
\,{f_{2}^2\,w_{1}\over w_{2}} \,X_{39}\,X_{47}\,X_{73}\,X_{94}
+f_{1}\,f_{3}\,X_{19}\,X_{57}\,X_{71}\,X_{95}
+f_{2}\,f_{3}\,X_{19}\,X_{57}\,X_{71}\,X_{95}\cr
&
+{f_{2}\,f_{3}\,w_{0}\over w_{2}} \,X_{39}\,X_{57}\,X_{73}\,X_{95}
+{f_{1}\,f_{3}\,w_{1}\over w_{2}} \,X_{39}\,X_{57}\,X_{73}\,X_{95}
-f_{2}^2\,X_{19}\,X_{67}\,X_{71}\,X_{96}\cr
&
+f_{1}\,f_{3}\,X_{19}\,X_{67}\,X_{71}\,X_{96}
-f_{2}^2\,X_{29}\,X_{67}\,X_{72}\,X_{96}
+f_{1}\,f_{3}\,X_{29}\,X_{67}\,X_{72}\,X_{96}\cr
&
-f_{2}^2\,X_{39}\,X_{67}\,X_{73}\,X_{96}
+f_{1}\,f_{3}\,X_{39}\,X_{67}\,X_{73}\,X_{96}
- {z_{0}\over z_{2}} X_{28} X_{47} X_{72} X_{84} \cr
&
+{f_{2}\,f_{3}\,z_{0}\over z_{2}} \,X_{29}\,X_{47}\,X_{72}\,X_{94}
+{f_{2}\,f_{3}\,z_{0}\over z_{2}} \,X_{29}\,X_{57}\,X_{72}\,X_{95}
-{z_{1}\over z_{2}} X_{28}\,\,X_{57}\,X_{72}\,X_{85} \cr
&+{f_{2}^2\,z_{1}\over z_{2}} \,X_{29}
\,X_{47}\,X_{72}\,X_{94}
+{f_{1}\,f_{3}\,z_{1}\over z_{2}}
\,X_{29}\,X_{57}\,X_{72}\,X_{95}  }
}
One can arrange the links
$X_{39},X_{29}$ and $X_{19}$ to correspond to $\ext^1$'s, and
by successively condensing them one should get exceptional collections,
quiver diagrams and
superpotentials for Del Pezzo 5,4, and 3 respectively.
There is no integrating out
involved so this is very simple. By doing an additional Seiberg
duality on node 7 in \higgsdpsix B, we get the quiver diagram for model IV yet
again and we have checked that one  recovers
the known superpotential for the quiver after going through this
sequence all the way.

The ${\bf C}^3/Z_3\times Z_3$  orbifold
is a limit of Del Pezzo 6 and its quiver diagram is given by \dpsix B. The
orbifold superpotential is completely cubic and our superpotential
should reproduce this in the appropriate limit. This looks promising
but we haven't been able to check it precisely due to the large number
of allowed field rescalings.

\subsec{Simple quiver diagrams for the remaining Del Pezzo's}

Let us list some quiver diagrams that can be deduced from
the collections in \Karpov\ for Del Pezzo surfaces of degree 1 and 2.\foot{
Quiver diagrams for low degree Del Pezzo's have been proposed before (see the fourth reference
in \FengBN) but they were found to have some
problems;  for degree 3 this was found by F.~Cachazo and the author, and for degrees 1 and 2
in \HerzogWT}
We haven't yet computed the
superpotentials for these quivers but they should give rise to cubic
couplings only.
%
%
\bigskip
\centerline{\epsfxsize=1.0\hsize\epsfbox{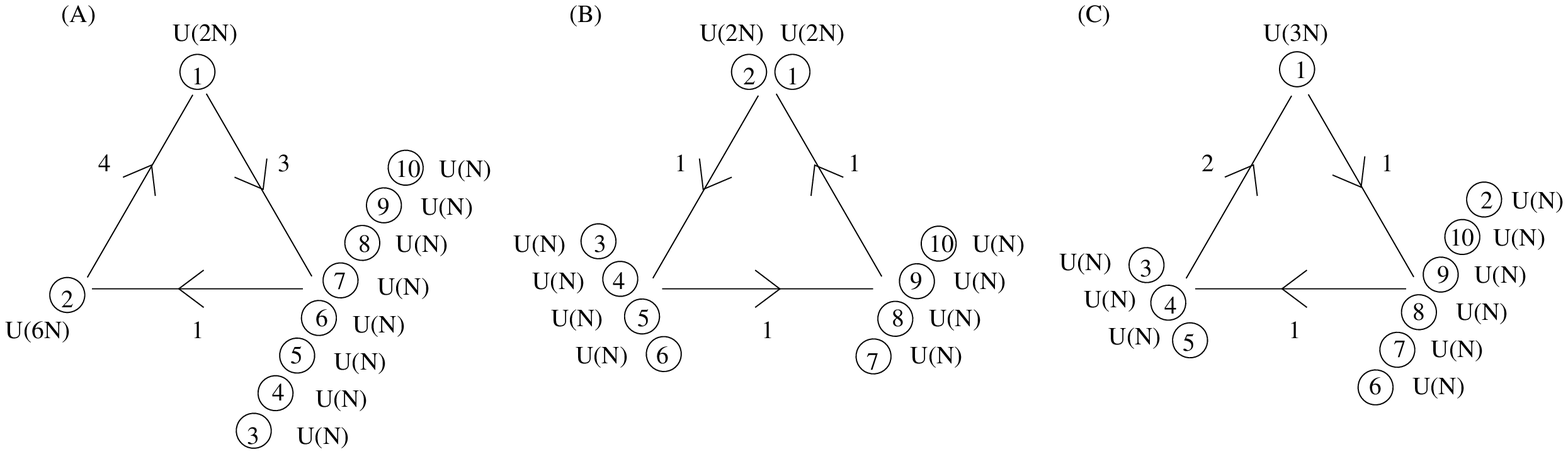}}
\nobreak\noindent{\ninepoint\sl \baselineskip=8pt
\fig\dpseven{}: {\sl
Three quivers diagrams for the degree 2 Del Pezzo.
All these should have a cubic superpotential.
(A): Type (7.1). (B): Type (7.2). (C): Type (7.3).}}
\bigskip


Let us start with collection (7.1) in \Karpov. The Chern
characters are computed to be:
\eqn\dpsevench{\matrix{
1.\ (2,-2 H + \sum_i E_i,-3/2) && 2.\ -(2,H,-1/2) && 3.\ (1,3H-\sum_i E_i,1) \cr
&&   && 4.\ (1,H-E_1,0) \cr
&&   && 5.\ (1,H-E_2,0)  \cr
&&   && 6.\ (1,H-E_3,0)  \cr
&&   && 7.\ (1,H-E_4,0)  \cr
&&   && 8.\ (1,H-E_5,0)  \cr
&&   && 9.\ (1,H-E_6,0)  \cr
&&   && 10.\ (1,H-E_7,0)  \cr
}}
The corresponding quiver diagram is drawn in figure \dpseven A.
The quiver can be simplified by dualising node 2, which reduces
the gauge group to $U(2N)$.

It is instructive to see how one may get the other allowed types of three-block
quivers through Seiberg duality. In principle we could get them from the collections in \Karpov,
 however computation of the cohomologies for the remaining collections
reveals that some of
them fail to be exceptional. In these cases
one may write down a valid set by
following the Chern characters through Seiberg dualities, as
we discussed in detail for the degree 5 Del Pezzo.

To get a new type of three-block quiver we can start with
\dpsevench\ and apply Seiberg dualities on nodes $\{2,3,4,5,6,2\}$
after which one arrives at the quiver depicted in \dpseven B. This
quiver belongs to the class of three-block collections called
(7.2) in \Karpov. Starting with \dpsevench\ and applying dualities
on nodes $\{2,3,4,5,1,2\}$ results in the quiver in \dpseven C,
which is of type (7.3).

Similarly we can get simple quiver diagrams for the degree 1 Del
Pezzo. Collection (8.1) in \Karpov\ appears to be invalid so we skip to
collection (8.2). The Chern characters are found to be
\eqn\dpeightch{\matrix{
1.\ (4,\sum_{i=1}^4 E_i,-5/2) && 2.\ -(2,H,-1/2) && 4.\ (1,H-E_1,0) \cr
&& 3.\ -(2,2H - \sum_{i=1}^3,-1/2) && 5.\ (1,H-E_2,0) \cr
&&   && 6.\ (1,H-E_3,0)  \cr
&&   && 7.\ (1,3H - \sum_i E_i + E_8,1)  \cr
&&   && 8.\ (1,3H - \sum_i E_i + E_7,1)  \cr
&&   && 9.\ (1,3H - \sum_i E_i + E_6,1)  \cr
&&   && 10.\ (1,3H - \sum_i E_i + E_5,1)  \cr
&&   && 11.\ (1,3H - \sum_i E_i + E_4,1)  \cr
}}
%
%
\bigskip
\centerline{\epsfxsize=.7\hsize\epsfbox{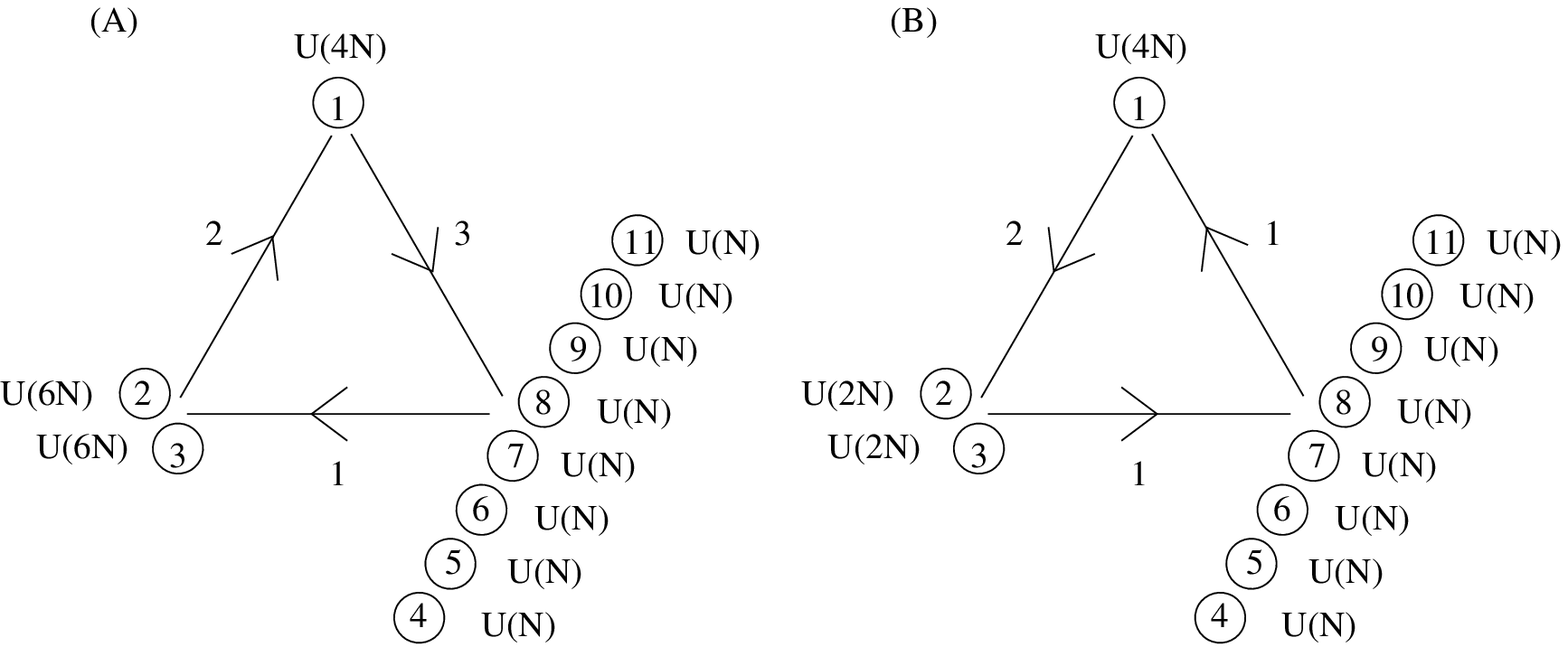}}
\nobreak\noindent{\ninepoint\sl \baselineskip=8pt
\fig\dpeight{}: {\sl
Quivers diagrams for the degree 1 Del Pezzo.
All these should have a cubic superpotential.
(A): Collection \dpeightch, type (8.2). (B): Seiberg dual.}}
\bigskip

%
The corresponding quiver diagram is shown in \dpeight A. A simpler quiver
may be obtained through Seiberg duality on nodes 2 and 3, drawn in
\dpeight B. From here we may obtain the other types of three-block
quivers through Seiberg duality. Type (8.1) can be recovered by
dualising $\{4,1,2,3\}$, type (8.3) by dualising $\{4,5,1\}$, and
type (8.4) by dualising $\{4,5,6,1,4,5,6\}$. The resulting
diagrams are given in \dpeightdual A, \dpeightdual B and \dpeightdual C respectively.
\bigskip
\centerline{\epsfxsize=1.0\hsize\epsfbox{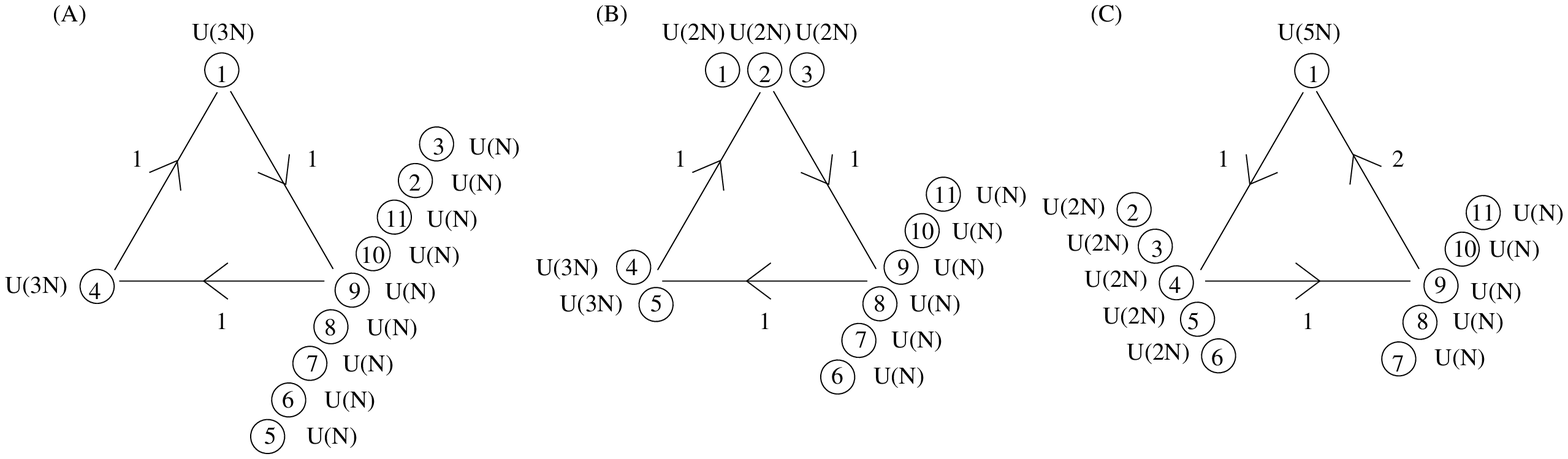}}
\nobreak\noindent{\ninepoint\sl \baselineskip=8pt
\fig\dpeightdual{}: {\sl
Other quivers diagrams for the degree 1 Del Pezzo
with cubic superpotential.
(A): Type (8.1). (B): Type (8.3). (C): Type (8.4). }}
\bigskip

%

\subsec{Preliminary remarks on the conifold}

In the case of the conifold the shrinking cycle is simply a ${ \bf
P}^1$ with normal bundle $N = \cO(-1) \oplus \cO(-1)$.
An exceptional collection is given by $\{ \cO(-1),\cO(0)
\}$. The non-zero cohomology groups are
\eqn\aa{  {\rm dim} \ \ext^0(\cO(-1),\cO(0)) = 2, \qquad
{\rm dim} \ \ext^0(\cO(-1), \cO(0)\otimes N) = 2  }
with the rest related by Serre duality. This gives rise to two
bifundamentals stretching one way between the two nodes and two  more
bifundamentals stretching the other way. The
first two fields describe the position of the D3-branes on the  ${ \bf
P}^1$ and the last two describe deformations in the normal
direction. It is hard to see how to compute the known quartic
superpotential, the current known rules do not suffice. Presumably one
wants to replace $\cO_{{ \bf P}^1}$ with some exact sequence of
branes/anti-branes  that fill the Calabi-Yau.

\subsec{$Z_k$ orbifolds}

While for orbifolds of the form ${\bf C}^3/Z_k$ the quiver diagram and
superpotential are easily derived using representation theory, we
would like to show here that the computations are not necessarily any
more difficult from a large volume perspective.

The (partial) resolution we would like to use for  ${\bf C}^3/Z_k$
is provided by the usual
linear sigma model. Suppose that the coordinates of  ${\bf C}^3$ are
labelled by $(x_0,x_1,x_2)$ and the weights of the action of $Z_k$ are
$(q_0,q_1,q_2)$ with $q_0+q_1 + q_2 = k$ and all $q$'s positive.
Then the  linear sigma model with four
fields given by  $ (x_0,x_1,x_2,p)$ and charges
$(q_0,q_1,q_2,-k)$ has a moduli space given by the solutions of
\eqn\aa{  q_0 |x_0|^2 +  q_1 |x_1|^2 + q_2 |x_2|^2 - k |p|^2 = t }
modulo the action of the $U(1)$. For large negative $t$ we get the
orbifold  ${\bf C}^3/Z_k$ and for large positive $t$ we get a partial
resolution with a finite size 4-cycle, namely the weighted projective space ${\bf
WP}(q_0,q_1,q_2)$ with the sheaf $\cO(-k)$ on top of it. In order to
derive the quiver theory we need a collection of sheaves on the
weighted projective space that correspond to the fractional branes for
the orbifold. It was argued in  \MayrAS\ that these should take the
form of  exterior powers of cotangent sheaves tensored with invertible
sheaves, and that the sections mapping between them should behave as
fermionic variables. One reason  we have to talk about sheaves and not bundles
is because weighted projective spaces typically have orbifold singularities.

We have not understood how to exactly identify the sheaves, but we
would like to point out that if sections behave as in
\MayrAS\ (see also \refs{\GovindarajanVI,\TomasielloYM})
then not only do we get the correct $Z_k$ symmetric quiver diagram, we also
get the correct superpotential.
Rather than setting up notation for the general case let us simply
treat an illustrative example. We would like to find the quiver for
the orbifold   ${\bf C}^3/Z_5$ with weights $(1,1,3)$. We will label
the sections by $\psi_0,\psi_1$ and $\psi_2$, where $\psi_i$ carries the
same $ U(1)$ charge as  $x_i$.
We denote the sheaves by $S_0, \ldots,
S_4$.
Then the claim is that $\ext^0(S_i,S_j)$ is generated by the fermionic
variables of total charge $j-i$.
So we obtain for example
\eqn\aa{ \eqalign{
{\rm dim }\ \ext^0(S_0,S_1) = 2 &\to
\psi_0,\psi_1 \cr
{\rm dim}\  \ext^0(S_0,S_2) = 1 &\to \psi_0 \psi_1 \cr
{\rm dim}\  \ext^0(S_0,S_3) = 1 &\to \psi_2 \cr
{\rm dim}\  \ext^0(S_0,S_4) = 2 &\to \psi_0 \psi_2, \psi_1 \psi_2 \cr
}}
Moreover if the fermion number of the map from $S_0$ to $S_i$ is odd,
then we have to invert the Chern character for $S_i$ in order to get
the correct quiver theory for D3 branes. The quiver diagram is drawn
in \wp.
\bigskip
\centerline{\epsfxsize=.50\hsize\epsfbox{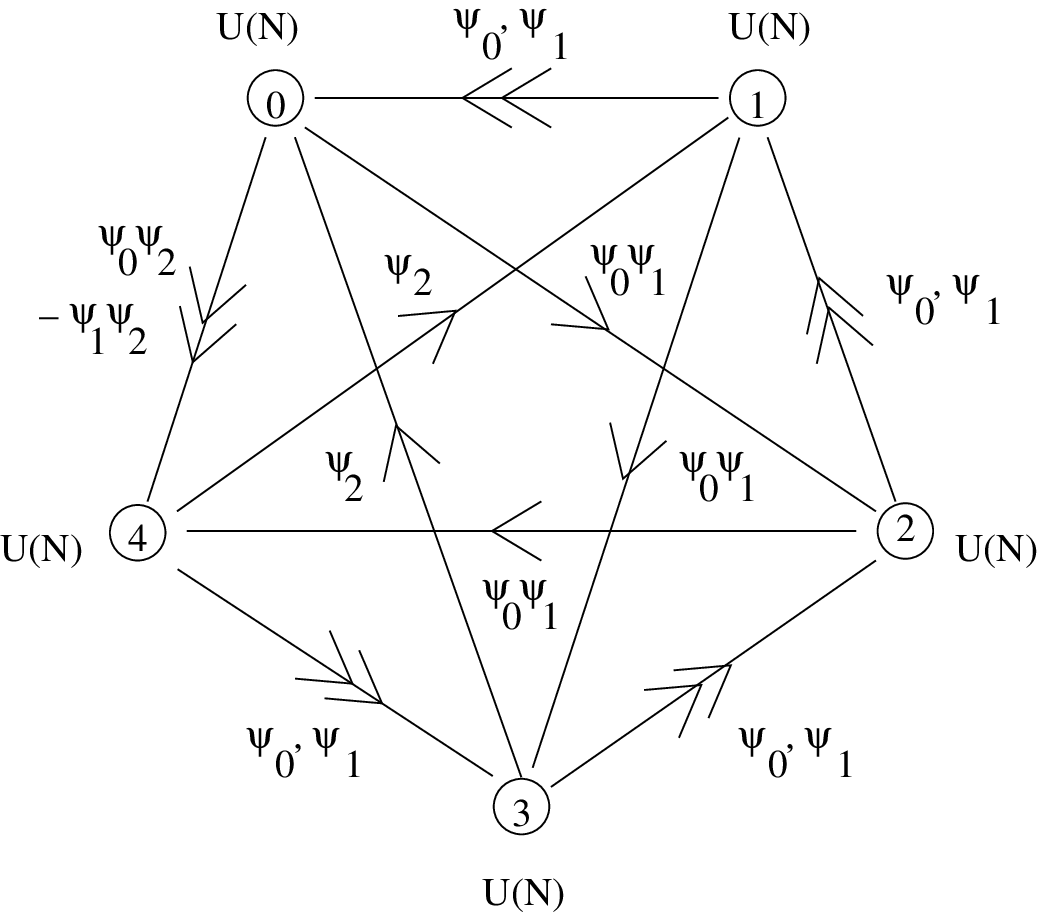}}
\noindent{\ninepoint\sl \baselineskip=8pt \fig\wp{} :
{\sl Quiver theory for the  ${\bf C}^3/Z_5$ orbifold.}}
\bigskip%
Now let us try to understand the superpotential. Because of the
fermionic nature of the variables we cannot multiply too many of them,
in fact we can only get cubic
couplings. Eg. composing the map $\psi_0$ from node 0 to node 1 with
the map $\psi_1$ from node 1 to node 2 yields a map $\psi_1 \psi_0$
from node 0 to node 2. Comparing with the standard generator which we
chose to be $\psi_0 \psi_1$, we see that we get a $-1$ as the
coefficient of the corresponding cubic term in the
superpotential. Continuing in
this way we find that
\eqn\aa{ \eqalign{
W&= (Y_{01} X_{12} - X_{01} Y_{12} ) Z_{20} +
 (Y_{12} X_{23} - X_{12} Y_{23} ) Z_{31} +
 (Y_{23} X_{34} - X_{23} Y_{34} ) Z_{42} \cr
&+
  (Y_{34} X_{40} - X_{34} Y_{40} ) Z_{03} +
 (Y_{40} X_{01} - X_{40} Y_{01} ) Z_{14}  }}
which is just the usual answer obtained by projecting the
superpotential of ${\cal N} = 4$ Yang-Mills theory.
The linear sigma model with charges
$(2,2,1)$ provides a different partial resolution of the same
orbifold, but leads to the same quiver diagram and superpotential up
to a permutation of the nodes.

\bigskip
\noindent
{\sl Acknowledgements:}

First and foremost I would like to thank F.~Cachazo, S.~Katz and C.~Vafa for
initial collaboration and many valuable discussions. In addition I  would like
to thank the organisers of the Workshop on Stacks and Computation at
Urbana-Champaign, June 2002, H.~Schenck and
M.~Stillman for introduction to and help with Macaulay2 at this
workshop,
Y.-H.~He for discussions on toric duality and
``chilling'' (i.e. doing M2 assignments at 2 a.m.), and the University
of Amsterdam, the Centre for Mathematical Sciences at  Zhejiang University, and the Morningside
Centre in Beijing for hospitality. This work was
supported in part by grant NSF-PHY/98-02709.

\listrefs

\bye